\documentclass[aps,prb,reprint,superscriptaddress]{revtex4-2}
\usepackage{amssymb, amsmath, amsfonts,color}
\usepackage{graphicx,xcolor}
\graphicspath{{./pics/}}
\usepackage{hyperref}
\usepackage{tikz}
\usetikzlibrary{braids}
\usepackage{bbold}
\usepackage{lipsum}
\usepackage{bm}
\usepackage{natbib}
\usepackage{placeins}
\setcitestyle{square}
\usepackage{braket}
\usepackage{simpler-wick}

\newcommand{\pd}{{\phantom{\dagger}}}

\begin{document}
\title{Fusion Dynamics of Majorana Zero Modes}

\author{Themba Hodge}
\affiliation{School of Physics, University of Melbourne, Parkville, VIC 3010, Australia}

\author{Tuan Kieu}
\affiliation{Department of Physics, University of Illinois at Chicago, Chicago, IL 60607, USA}

\author{Jasmin Bedow}
\affiliation{Department of Physics, University of Illinois at Chicago, Chicago, IL 60607, USA}

\author{Eric Mascot}
\affiliation{School of Physics, University of Melbourne, Parkville, VIC 3010, Australia}

\author{Dirk K. Morr}
\affiliation{Department of Physics, University of Illinois at Chicago, Chicago, IL 60607, USA}

\author{Stephan Rachel}
\affiliation{School of Physics, University of Melbourne, Parkville, VIC 3010, Australia}

\begin{abstract}
Braiding and fusion of Majorana zero modes are key elements of any future topological Majorana-based quantum computer. Here, we investigate the fusion dynamics of Majorana zero modes in the spinless Kitaev model, as well as in a spinful model describing magnet-superconductor hybrid structures. We consider various scenarios allowing us to reproduce the fusion rules of the Ising anyon model. Particular emphasis is given to the charge of the fermion obtained after fusing two Majorana zero modes: as long as it remains on the superconductor, charge quantization is absent. When moving the fermion to a non-superconducting region, such as a quantum dot, nearly-quantized charge can be measured. Our findings confirm for both platforms that fusion dynamics of Majorana zero modes can indeed be used for the readout of Majorana qubits.
\end{abstract}

\maketitle
%
Majorana zero modes (MZMs), which emerge as zero-energy in-gap states in topological superconductors, are a promising candidate for the implementation of fault-tolerant quantum computation\,\cite{Nayak2008, Beenakker2020, Pachos2012} based on their non-Abelian braiding statistics. Numerous proposals have been made on how topological quantum gates can be realized in both nanowires and magnet-superconductor hybrid (MSH) structures\,\cite{Alicea2011, Cheng2011, Amorim2015, Sekania2017, Sanno2021, Mascot2023, Bedow2024, Hodge2024}.
To test the non-Abelian braiding properties of MZMs and the success of any topological quantum gate in either structure, the ability to read-out the state of a Majorana qubit is essential. For this purpose, MZMs have to be fused, i.e., a pair of MZMs is moved onto a single position. This process is governed by the fusion rules of Ising anyon model. The outcome of this fusion process then needs to be detected, for which the usage of quantum dots has been suggested\,\cite{Smith2020, Steiner2020, Zhou2022, Sau2024, Tsintzis2024}.

The fusion rules lay out how the different possible ``objects'' behave under fusion. These objects are conventional complex fermions, i.e., superconducting quasiparticles and anyons, i.e., the MZMs. In addition, there is the {\it vacuum} which can be considered as the neutral element for the fusion rules; practically, one can think of it as the superconducting condensate. The non-Abelian character of the MZMs is encoded in the fusion rule of two MZMs with each other, which has two different channels: either the two MZMs are fused into the vacuum, i.e., absence of any particle; or they are fused into a complex fermion, i.e., a charge can be absorbed. Explicating and investigating these abstract, mathematical rules in a physical and time-dependent system is the main aim of this paper.

Previous works have explored static aspects of fusing MZMs, including Refs.\,\cite{Bai2024, Tsintzis2024, Sau2024, Steiner2020}.
In contrast, work involving not only the dynamical properties of fusion in a full many-body context but also how multiple braids affect the final fusion result in a full many-body simulation, critical for a universal readout scheme, is mostly absent from the literature.
Such investigations on different models is critical in the context of topological quantum computation, as the effects of \emph{diabatic error}, \emph{braiding error} and \emph{charge quantization} at the fusion site, which all require full, many-body simulations to probe. 
An interesting exception constitutes the paper by Boross et al.\,\cite{Boross2024} where on a Kitaev chain, different initial states are fused after a routine involving both a braid, and a subsequent phase gate, to institute and readout an effective X-gate in a full dynamical context.

In this paper, we first give an overview of the fusion rules of MZMs\,\cite{Verlinde1988, Lahtinen2017, Maciazek2024}, which are those of Ising anyons. While some proposals have been made for implementing these fusion rules\,\cite{Souto2022, Pandey2023, Pandey2023_2,Pandey2024, Wang2024}, we extend these approaches here, using recent advances in simulating time-dependent phenomena in superconductors without an exponentially scaling Hilbert space. We implement the corresponding fusion processes in both the Kitaev model for nanowires\,\cite{Kitaev2001, Kitaev2006} and a spinful model for MSH structures\,\cite{Nadjperge2013,Braunecker2013,Pientka2013,Klinovaja2013,Ruby2015,Pawlak2016}, showing that both models are suitable for moving, braiding, and fusing MZMs with read-out, tractable charges to confirm the predictions of the fusion rules. We study Z-fusion in both models, by fusing two MZMs from the same pair, where we demonstrate that the outcome of this process can detect the parity of a single pair of MZMs and show for both the Kitaev and the MSH model how the fusion charge can be localized, even after the completion of the fusion.

Moreover, we also consider fusion in the X-channel, where two MZMs from different pairs are fused, in the Kitaev model, thus demonstrating the non-trivial fusion channel which gives rise to the non-Abelian fusion statistics inherent with the Ising anyon model.
Finally, we expand on previous work by studying fusion processes after completion of different quantum gates, such as the $\sqrt{\rm{X}}$, X and Hadamard gates. The implementation of these routines demonstrate the critical role braiding has, not only in changing the quantum state of an initial qubit, but also, along with utilizing different fusion bases, how it can be used, along with fusion, to gain full state information for a single logical qubit on the Bloch sphere.

The paper is organized as follows: In Sec.\,\ref{sec:kitaev} we focus on the spinless Kitaev model, and in Sec.\,\ref{sec:MSH} on the spinful model describing MSH structures. In both sections we discuss various aspects of fusion. In particular, we consider fusion after performing braids corresponding to single-qubit gates. In Sec.\,\ref{sec:conclusion} we present a conclusion.

%
\section{Fusion on Kitaev Wires and networks}\label{sec:kitaev}
%

\subsection{Fusion of MZMs}
For $\textbf{1}$ (vacuum), $\sigma$ (anyon) and $\zeta$ (fermion), the Ising anyon model satisfies
\begin{equation}
\begin{split}
     & \sigma\times \sigma=\textbf{1}+\zeta, \quad
    \sigma\times \zeta=\sigma, \quad
     \zeta\times \zeta=1\\
     & \textbf{1} \times \textbf{1}=\textbf{1}, \quad \quad  \: \; \; \textbf{1}\times \sigma=\sigma,\quad \textbf{1}\times \zeta=\zeta. 
    \end{split} \label{eq:fuserules}
\end{equation}
In the context of a topological superconductor, such as a chiral $p$-wave superconductor, we can not only interpret the fusion rule $\zeta \times \zeta=1$ as the condensation of two fermions into the Cooper sea, but we can also identify the anyon $\sigma$ with an MZM\,\cite{Read2000,Ivanov2001}.
For the purposes of topological quantum computation, the most important of these rules is the two-dimensional fusion channel $\sigma \times \sigma=\textbf{1}+\zeta$. 
We associate each anyon with a \emph{topological charge}, with $\textbf{1}$, $\sigma$, $\zeta$ having $0$, $\frac{1}{2}$, and $1$, respectively. 
Over each fusion channel, the total topological charge \emph{must} be conserved\,\cite{Nayak2008} and 1 is the maximum possible charge of a channel (e.g. $\sigma \times \zeta \equiv \frac{1}{2}\times 1=\frac{1}{2}$ rather than $\sigma \times \zeta \equiv \frac{1}{2}\times 1=\frac{1}{2},\frac{3}{2}$). 
This encoding provides the model with its non-Abelian character, as changing the fusion order of four or more anyons may change the result of the final fusion outcomes.

\begin{figure}[t!]
    \centering
    \includegraphics{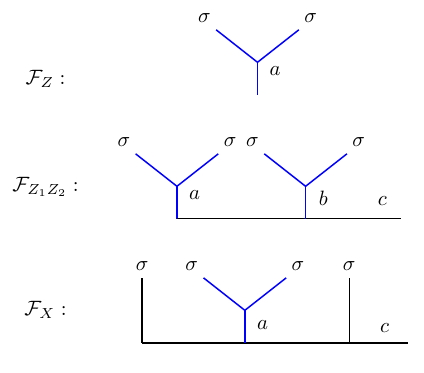}
    \caption{Schematic of Fusion processes.
    (a) $\mathcal{F}_Z$ fusion process, which corresponds to the fusion of the two MZMs that make up a Majorana bound state. 
    (b) $\mathcal{F}_{Z_1Z_2}$ fusion process, where in a system with two Majorana bound states (consisting of four MZMs), the two MZMs on the left and the two MZMs on the right are pairwise fused.
    (c) $\mathcal{F}_X$ fusion process where we fuse the middle two MZMs.
    The topological charges $a,b,c\in \{1,\zeta\}$ in all cases, with $c$ setting the total topological charge in (b) and (c). In the dynamical simulations, we are interested in determining $a$ or $a$ and $b$.}
    \label{fig:Fig0Kit}
\end{figure}

    The two relevant fusion orders considered in this paper are referred to as $Z$ and $X$ fusion, in the following denoted by $\mathcal{F}_{Z}$ and $\mathcal{F}_X$. For a system consisting of only two MZMs, there is only one possibility of fusion, $\mathcal{F}_{Z}$, as depicted in Fig.~\ref{fig:Fig0Kit}\,(a). For two pairs of MZMs, there are obviously more options: One can either perform a $Z$ fusion for each pair individually, denoted as $\mathcal{F}_{Z_1 Z_2}$ and depicted in Fig.~\ref{fig:Fig0Kit}\,(b). Alternatively, one can fuse  two MZMs belonging to the different pairs, i.e., topological segments of the system; the latter is refered to as $\mathcal{F}_X$ and depicted in Fig.~\ref{fig:Fig0Kit}\,(c). We note that the considered $\mathcal{F}_X$ process leaves the two outer MZMs unfused.

     Each pair of MZMs corresponds to a single physical qubit, and we will label the corresponding qubit states as $\ket{0}$ and $\ket{1}$ for simplicity. The former has even parity, while the latter has odd parity, corresponding to the formation of an electron after fusion, with both states differ by one charge $-e$. The manifold $\{\ket{0}, \ket{1}\}$ is sometimes also referred to as {\it fusion space}.
     For a system consisting of two pairs of MZMs, i.e., two physical qubits, we label them accordingly as $\ket{00}$ and $\ket{11}$ (even parity) and as $\ket{01}$ and $\ket{10}$ (odd parity). In this case, the corresponding fusion space contains all the four states.

     A single physical qubit cannot be used for practical purposes, as the states $\ket{0}$ and $\ket{1}$ differ by one charge and thus by their parity. A coherent superposition $\ket{0}\pm\ket{1}$ is not possible. Instead, one constructs a logical qubit as a pair of physical qubits, where the second qubit plays the role of an auxiliary qubit to keep the total parity conserved. Throughout the paper, we will label logical qubit states as $\ket{\,\cdot\,}_{\rm logic}$. In the even parity sector, we identify $\ket{00}\to \ket{0}_{\rm logic}$ and $\ket{11}\to\ket{1}_{\rm logic}$. In the odd parity sector, we identify $\ket{01}\to \ket{0}_{\rm logic}$ and $\ket{10}\to\ket{1}_{\rm logic}$.

    For generic physical or logical single-qubit states, it will be convenient to define them on the Bloch sphere as usual, parametrized by the two angles $\theta$ and $\phi$, $\ket{\psi(\theta,\phi)} = \left( \cos{\frac{\theta}{2}}\ket{0} + e^{i\phi}\sin{\frac{\theta}{2}}\ket{1}\right)$.

    $\ket{0}_{\rm logic}$ and $\ket{1}_{\rm logic}$ are the canonical states for the $Z$ fusion basis, but as usual we could also define physical qubit states in the $X$ fusion basis, $\ket{\pm}_{\rm logic} = \tfrac{1}{\sqrt{2}}(\ket{0}_{\rm logic}\pm\ket{1}_{\rm logic})$.

     In this paper, we will consider both $\mathcal{F}_{Z}$ and $\mathcal{F}_X$. Conveniently, the Ising anyon model allows to change the fusion order by a linear transformation $F$, the so-called $F$-move, which in case of the non-Abelian fusion rule, $\sigma\times\sigma=\mathbf{1}+\zeta$, appears to be a basis change.
    One might, hence, not be surprised that the matrix representation of this $F$-move is none other than a Hadamard gate. Details about the $F$-move are delegated to the Appendix.
    Further, as well known in the case of MZMs, we may also change the fusion outcome by exchanging or \emph{braiding} the anyons prior to fusion, which will lead to a unitary transformation on the Majorana ground-state manifold. 

     In order to verify fusion predictions, we need to measure (local) charges and differences of charge configuration. In particular, we will measure charge differences with respect to a {\it reference} state. While this might in principle be a random state, in this paper we restrict ourselve to the following reference states (unless explicitly mentioned otherwise): For $\mathcal{F}_{Z}$ we choose $\ket{0}$ and for 
        $\mathcal{F}_{X}$ we choose $\tfrac{1}{\sqrt{2}}\big(\ket{00}+\ket{11}\big) \equiv \ket{+}_{\rm logic}$.  
        For the process $\mathcal{F}_{Z_1 Z_2}^{\rm even}$ our reference state is  $\ket{00}$, but for $\mathcal{F}_{Z_1 Z_2}^{\rm odd}$ $\ket{01}$. For $\mathcal{F}_{Y}$ we use $\tfrac{1}{\sqrt{2}}\big(\ket{00}+i \ket{11}\big)$ as reference state.
        
     For an arbitrary state $\ket{\psi(t)}$, we defined the local charge or charge density
    \begin{equation} \rho_i(t) \equiv \rho(\vec r, t) = -e \bra{\psi(t)} n_i \ket{\psi(t)}\label{eq:localcharge}
    \end{equation}
    with the number operator $n_i=c_i^\dag c_i^\pd$.
    The local charge of the reference states is denoted as $ \rho^{\rm ref}_i(t)$. The total charge is then defined as
    \begin{equation}\label{eq:totalcharge}
    Q(t) = \sum_i \rho_i(t)
    \end{equation}
    and as $Q_{\rm ref}(t)= \sum_i \rho^{\rm ref}_i(t)$. Finally, we denote the total charge of an arbitrary single qubit state as $Q(\theta,\phi)(t)$.

     The central object of this paper is the charge difference a state after fusion and its reference state, denoted as $\Delta Q(t)$. We define it as
    \begin{equation}
        \label{eq:Qdiff}
        \Delta Q(t) = Q(t) - Q_{\rm ref}(t)\ .
    \end{equation}
    Accordingly, we will also use $\Delta Q(\theta,\phi)(t)$.
    
     As discussed later in the text, each fusion result corresponds to a measurement of Pauli $\langle\sigma_z\rangle=\textrm{cos}(\theta)$ and $\langle\sigma_x\rangle=\textrm{sin}(\theta)\textrm{cos}(\phi)$ for an $\mathcal{F}_{Z_1Z_2}$ and $\mathcal{F}_X$ routine, respectively, providing both population and phase information for the state.
   
     We note, however, that all readouts in the paper are given in terms of the expectation values of a local charge density, $\rho_i(t)$. 
    These readouts are probabilistic. In terms of an experimental readout, projective measurements routines, such as the proposed charge capacitance routine \cite{Zhou2022}, that would collapse the fusion result to $\textbf{1}$ or $\zeta$ after measurement. After many measurements, we expect the averaged readouts to converge to the correct result. 
    In this way, we are effectively probing the ensemble outcome of these measurements, which effectively reveals the quantum information of the many-body state.
    It is worth noting that single-fusion results do have a vital role in quantum computing, such as measurement based architectures\,\cite{Bonderson2008,Briegel2009,Karzig2017}. 
    Dynamical analysis of such methods in the case of MZM systems we leave for future work.

%
\subsection{Model} 
%
We first consider the simplest model for an emergent topological superconductor, the spinless, one-dimensional $p$-wave superconductor\,\cite{Kitaev2001}. 
The Hamiltonian, $\mathcal{H}_{\rm Kit}$, for this system is given by

\begin{align}
\mathcal{H}_{\rm Kit} = -\mu \sum_r c_r^\dagger c_r^\pd
- \sum_r \left(
    \tilde{t} c_r^\dagger c_{r+1}^\pd + 
    \Delta c_r^\dagger c_{r+1}^\dagger + {\rm H.c.}
\right),
\end{align}
where $\mu$ is the chemical potential, $\tilde{t}$ is the hopping amplitude, and $\Delta = |\Delta| e^{i\phi}$ is a phase dependent $p$-wave pairing amplitude.
The topological regime is realized in the regime where $-2<\mu_{\textrm{topo}}<2$, with MZMs emergent on the boundary of the material.

We may move the MZMs by utilizing a time-dependent chemical potential along each site, $\mu_i(t)$, which ramps the chain between a trivial regime and a topological regime, effectively shifting the topological boundary of the material. 
This may be encoded by 
utilizing a smooth-step polynomial function, $s$, which interpolates the chemical potential, $\mu$, between some trivial choice, $\mu_{\textrm{triv}}$, and some topological choice, $\mu_{\textrm{topo}}$. 
For $0 \leq q \leq 1$, where $q=0$ denotes the beginning of the step, and $q=1$ the end of the step, this function is given by

\begin{align}
    s(q) = \begin{cases}
    0 & q \leq 0, \\
    q^2 (3-2q) & 0 \leq q \leq 1, \\
    1 & 1 \leq q
    \end{cases}.
\end{align}

Using this, we set the time-dependent chemical potential along each lattice point $i$ to be

\begin{equation}
    \mu_i (t) = \mu_{\rm triv} + (\mu_{\rm topo} - \mu_{\rm triv}) \,\,s\left( \frac{t}{\tau}(1 + \alpha (N-1)) - \alpha i \right)\ .\label{eq:ramp}
\end{equation}
Here $\alpha$ is a \emph{delay coefficient} that scales the delay between the start of the ramp along each site on the leg\,\cite{Sekania2017,Mascot2023,Hodge2024,Peeters2024}.
This provides an interpolation between two critical regimes, whereby the MZM is smoothly dispersed along the chain as the chemical potential of each site is ramped simultaneously ($\alpha=0$), and the regime where the MZM is ramped site-by site ($\alpha=1$). 
The delay coefficient is set to $\alpha=0.025$ throughout this work unless otherwise specified.

%
\subsection{Time-Evolution} 
%

We look to dynamically calculate the charge as a function of time. 
We do this by not only considering the evolution in the MZM subspace, but under the influence of the entire many-body Hilbert space, whereby non-adiabatic contributions such as quasiparticle poisoning  \cite{Cheng2011,Karzig2021} may lead to error in the fusion result. 

We first consider the Fock state, $|\vec{n}_d\rangle=\prod_k (d^{\dag}_k)^{n_k}|0_d\rangle$, with $n_k \in \{0,1\}$.
This corresponds to the filling of excited states, with associated single particle energies $E_k\geq 0$.
As such, $|0_d\rangle = \prod_k d_k|0_c\rangle$ will correspond to the ground state of the Hilbert space, where $|0_c\rangle$ is the fermionic vacuum, defined such that $c_i|0_c\rangle=0$ $\forall i$. 
In this way, the ground state, $|0_d\rangle$, corresponds to the lowest energy many-body state within the fusion space\,\cite{Boross2024}, with excitations restricted to the low-energy subspace filling the rest of the fusion space, and higher energy excitations corresponding to bulk states.
For $\mathcal{H}(t)$ a time-dependent Hamiltonian, the time-evolution $\mathcal{S}(t)$ may be expressed as the time-ordered integral $\mathcal{S}(t)=\mathcal{T}\textrm{exp}(-i\int^t_{0}\mathcal{H}(t')dt')$. 
As such, we may express the time-evolved Fock state as
\begin{equation}
    |\vec{n}_d(t)\rangle=\mathcal{S}(t)\prod_k(d^{\dag}_k)^{n_k}|0\rangle_d=\prod_k(d^{\dag}_k(t))^{n_k}|0(t)\rangle_d.
    \label{eq:Time-evolution}
\end{equation}
The quasiparticle operators $d^{\dag}_k(t)$ may be found utilizing the \emph{time-dependent Bogoliubov equations}, where for $c_i=U_{ij}d_j+V_{ij}d^{\dag}_j$, the time-evolved quasiparticles are similarly connected to the quasiparticle basis by $c_i=U_{ij}(t)d_j(t)+V_{ij}(t)d^{\dag}_j(t)$, with time-evolution given by
\begin{equation}
    i\partial_t\begin{pmatrix}
    U(t) & V(t)^* \\ 
    V(t)  & U(t)^*
    \end{pmatrix}=H_{\textrm{BdG}}(t)\begin{pmatrix}
    U(t) & V(t)^* \\ 
    V(t)  & U(t)^*
    \end{pmatrix}. 
\end{equation}
The matrix $H_{\rm BdG}(t)$ is the time-dependent Bogoliubov--de Gennes (BdG) Hamiltonian for the model, where $\mathcal{H}(t)=\frac{1}{2}\Psi^{\dag}(t)H_{\rm{BdG}}(t)\Psi(t)$, with $\Psi(t)=\left(c_1(t), \ ...,\, c_N(t), \ c^{\dag}_1(t),\ ..., \ c^{\dag}_N(t)\right)^T$. 
We then employ the \emph{time-dependent Pfaffian Method} introduced in \cite{Mascot2023}.  
By mapping $\{d_k\}$ to a canonical basis $\{\bar{d}_k\}$ by the unitary transformation $d_i=D_{ij}\bar{d}_j$, with the unitary $D_{ij}$ gained from a singular value decomposition of $U$. 
We may now transform the Bogoliubov vacuum as $|0_d\rangle=\frac{1}{\sqrt{N}} \prod_{k \in P} \bar{d}_k \bar{d}_{\bar{k}} \prod_{k \in O} \bar{d}_k \ket{0_c}$, with $N$ a normalisation factor, with $P$ and $O$ indexing the paired and occupied indicies of the decomposition \cite{Mascot2023}. 
Thus, by an application of Wick's theorem, we may calculate the local charge density at any site $i$ by considering the correlation function between two arbitrary Fock states $\mathcal{N}_{mn}(t)=\langle \vec{n}_{d_{m}}(t)|n_i|\vec{n}_{d_{n}}(t)\rangle$. 
This may be calculated by 
\begin{equation}
\begin{split}
    &\mathcal{N}_{mn}(t)=\;  \pm\frac{ 1}{N(t)}\textrm{pf}\\
    &\begin{pmatrix}
        \wick{\c {\bar{d}}^\dag(t) \c {\bar{d}}^\dag(t)} &
        \wick{\c {\bar{d}}^\dag(t) \c d_m(t)}&\wick{\c {\bar{d}}^\dag(t) \c {n_i}}&
        \wick{\c {\bar{d}}^\dag(t) \c d_n^\dag(t)} &
        \wick{\c {\bar{d}}^\dag(t) \c {\bar{d}}(t)} \\[3pt]
        & 
        \wick{\c d_m(t) \c d_m(t)} & \wick{\c d_m(t) \c {n_i}}&
        \wick{\c d_m(t) \c d_n^\dag(t)} &
        \wick{\c d_m(t) \c {\bar{d}}(t)} \\[3pt]
         && \wick{\c {n_i} \c {n_i}}&
        \wick{\c {n_i} \c d_n^\dag(t)} &
        \wick{\c {n_i} \c{\bar{d}}(t)}
       \\[3pt] 
        &&&
        \wick{\c d_n^\dag(t) \c d_n^\dag(t)} &
        \wick{\c d_n^\dag(t) \c {\bar{d}}(t)} \\[3pt]
        &&&&
        \wick{\c {\bar{d}}(t) \c {\bar{d}}(t)}
    \end{pmatrix}
\end{split}
\end{equation}
where the additional matrix elements are found by the antisymmetry of the correlation matrix and ${\rm pf}(\cdot)$ is the Pfaffian of a matrix.
Further, the $\pm = (-1)^{n_{\bar{d}}(n_{\bar{d}}-1)/2 + n_d(n_d-1)/2}$ 
where $n_{\bar{d}}$ [$n_d$] correspond to the number of $\bar{d}^\dag(t)$ [$d(t)$] operators in $\bra{\vec{n}_{d_m}(t)}$.
Here, each contraction corresponds to a matrix of the vacuum expectation values $\wick{\c {a} \c {b}}=\langle 0_c| ab|0_c\rangle$. 

As such, for an arbitrary state, $|\psi\rangle=\sum_ka_k|\vec{n}_{d_{k}}\rangle$, the local charge density may then be calculated by calculating the sum $\rho_i(t)=-e\sum_{kl}a^*_ka_l\mathcal{N}_{kl}(t)$.
The total charge along the chain at time $t$ may now be calculated by $Q(T)=\sum_i\rho_i(T)$, providing a measure for the total charge after fusion.
This provides a measure for the fusion that considers not only the physics within the Majorana subspace, but also the rest of the Hilbert space, all of which will influence the final quantum information we retain through the process.

\begin{figure}[t!]
    \centering
    \includegraphics{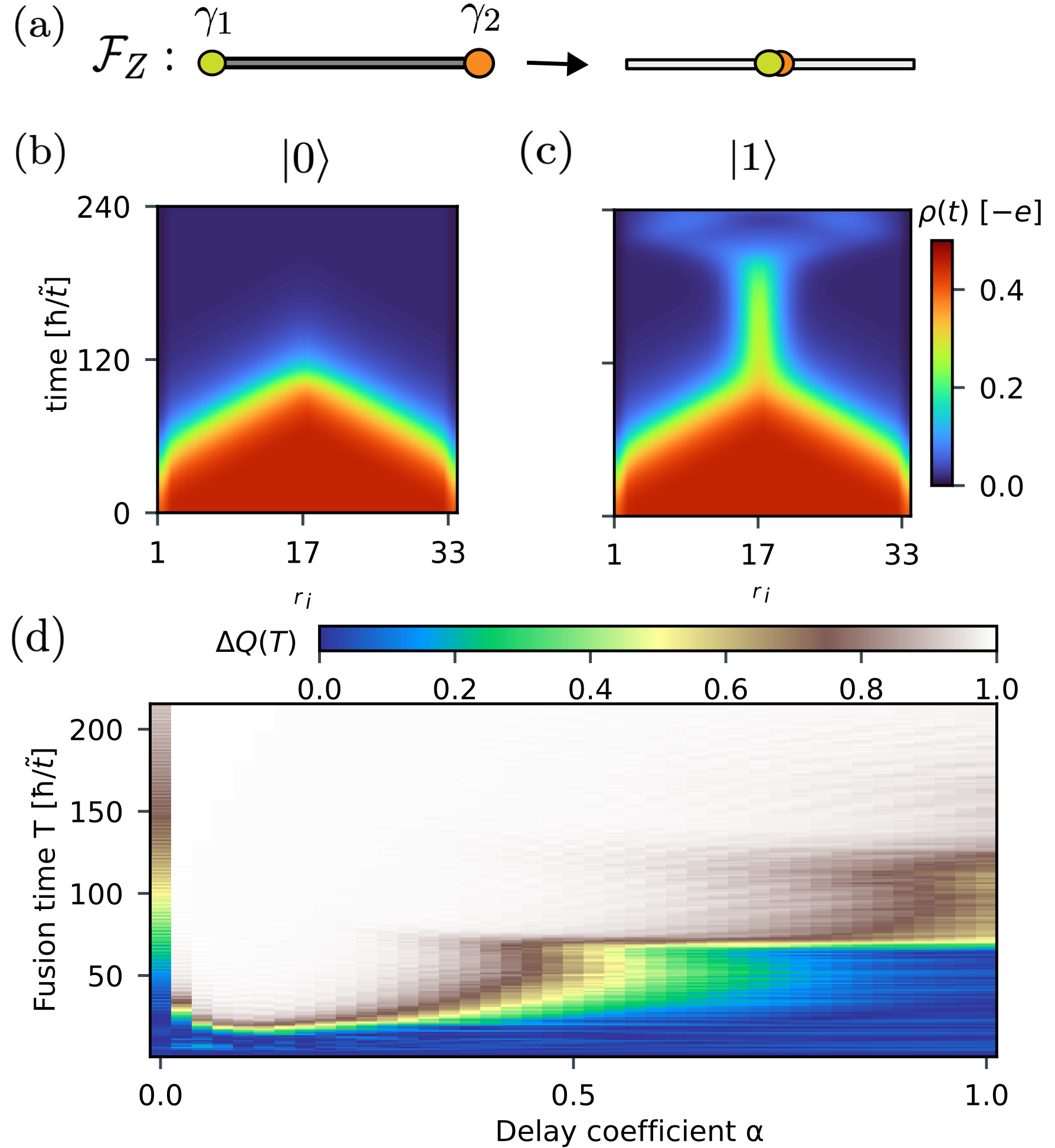}
    \caption{Fusion through the $Z$-channel on a Kitaev wire. (a) Graphic of the $\mathcal{F}_Z$ process on a single Kitaev chain. (b), (c) Charge density $\rho_i(t)$ for fusion of two MZMs, initialized at $|0\rangle$ and $|1\rangle$, respectively. (d) The final charge difference $\Delta Q(T)$ between the $|1\rangle$ and $|0\rangle$ states as a function of fusion time $T$ and delay coefficient $\alpha$ of the ramping process. Parameters used for these simulations are $(\tilde{t},\Delta,\mu_{\rm topo}, \mu_{\rm triv},L)=(1,1,-0.4,-5.6,32a_0)$.}
    \label{fig:Fig1Kit}
\end{figure}

%
\subsection{$Z$-Fusion}\label{sec:Z-fusion}
%

We begin by considering the simplest scenario, a pairwise fusion between an initialized pair of MZMs $\gamma_1$ and $\gamma_2$.
As shown in Fig.\,\ref{fig:Fig1Kit}\,(a), we do this by ramping a single topological region to trivial. We start at the chain ends, ramping from the topological regime to the trivial regime, until the MZMs have been transported to the middle of the wire, leading to the forming of a massive fermion or quasiparticle state, as the MZM wavefunctions hybridize over the approach. 

We first consider the pure states $|0\rangle$ and $|1\rangle$. 
As we are dealing with only one pair of MZMs, we will denote this as an $\mathcal{F}_Z$ fusion process.
As the occupancy of the state encodes its topological charge, we expect the calculated $\Delta Q(T)=-en$ for the single fusion event, where $n =0$ $(1)$ for the $|0\rangle$ $(|1\rangle)$ states, respectively.
This is shown in Fig.\,\ref{fig:Fig1Kit}\,(b) and (c), where we plot the charge density, $\rho_i(t)$, along the wire as a function of time. 
For the unfilled many-body state, $|0\rangle$, shown in Fig.\,\ref{fig:Fig1Kit}\,(b), we see no signal at the fusion site in the middle of the chain, with the charge density diminishing, $\rho_i\to 0$, as the chemical potential sits below the dispersion. 
However, if we set $|\vec{n}_d\rangle=|1\rangle$, a clear signal forms at the fusion site, as each MZM is ramped towards each other (see Fig.\,\ref{fig:Fig1Kit}\,(c)). 
This corresponds to the forming of a massive fermion, confirmation of the Ising anyon fusion statistics. 
Further, as we complete the ramping routine at $T=200 \hbar/\tilde{t}$, the formation of a lightcone in the charge density indicates that the fermion is no longer pinned to the site, and disperses into the wire freely. 

In Fig.\,\ref{fig:Fig1Kit}\,(d), we test quantization of the fusion by simulating the result for varying fusion times $T$ and delay coefficient $\alpha$. 
In doing so, we are able to examine how adiabatic effects adjust the fusion result. 
To measure the associated charge of fusion, we calculate the charge difference, $\Delta Q$, between the two initializations $|1\rangle$ and $|0\rangle$. 
Positively, we can see, for longer fusion times, corresponding to a more adiabatic ramping routine, we find total charge difference after the fusion result approaches $-e$, with maximum charge difference found to be $\Delta Q(T)=-0.996e$ at $(T,\alpha)=(216\hbar/\tilde{t},0.025)$ over the parameter space. 
While the result diminishes as we approach a site-by-site ramping routine as $\alpha \to 1$, in general, we still find the result improves as a function of braid time.
This suggests not only do we get the expected fusion result, pivotal for a functioning readout routine, but further, the result is improved as a function of braid time, with convergence towards $-e$ in all cases as we enter the adiabatic regime.

%
\subsection{$X$-Fusion} \label{sub:X}
%
\begin{figure}[t!]
    \centering
    \includegraphics{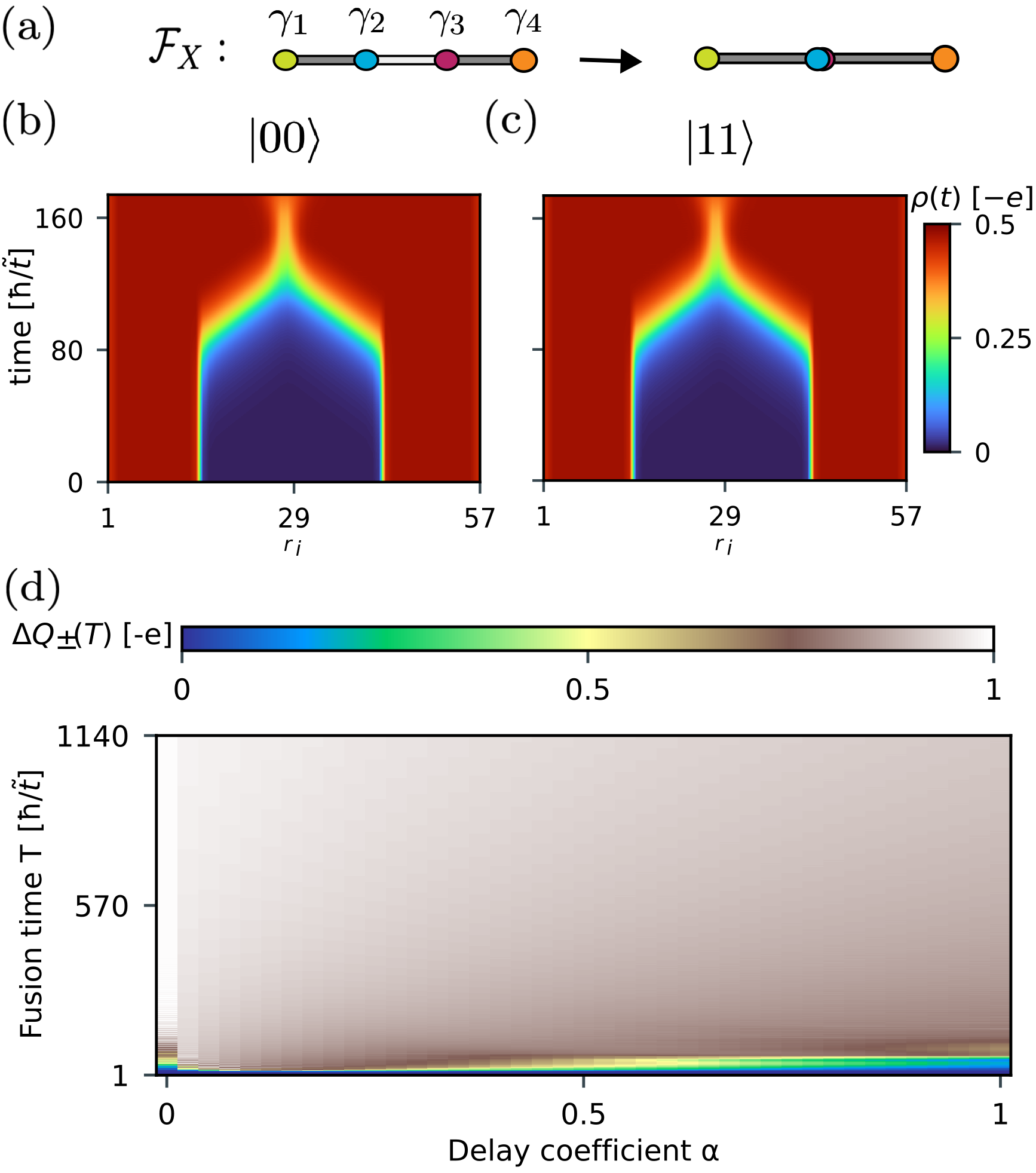}
    \caption{Fusion through the $X$-channel on a Kitaev wire. (a) Graphic of the $\mathcal{F}_X$ process on a single Kitaev chain for two pairs of MZMs. (b), (c) Charge density $\rho_i(t)$ for $X$-channel fusion between two pairs of MZMs, initialized in the even parity sector with $|00\rangle$ and $|11\rangle$, respectively. (d) The final charge difference, $\Delta Q(T)$, between the state $\ket{-}$ and the reference state $\ket{+}$ as function of fusion time $T$ and delay coefficient $\alpha$. Parameters used for these simulations are $(\tilde{t},\Delta,\mu_{\rm topo}, \mu_{\rm triv},L)=(1,1,-0.45,-4.4,56a_0)$.}
    \label{fig:Fig2Kit}
\end{figure}
We next consider fusion between MZMs belonging to different Majorana bound states. 
As shown in Fig.~\ref{fig:Fig2Kit}\,(a), we can encode this on a Kitaev wire by considering two separate topological regions, with $\gamma_1$ ($\gamma_3$) and $\gamma_2$ ($\gamma_4$) arising on the boundaries of the left (right) region. 
Each Majorana bound state is now given by the quasiparticle operator $d^{\dagger}_1=\frac{1}{2}\big(\gamma_1-i\gamma_2\big)$, $d^{\dagger}_2=\frac{1}{2}\big(\gamma_3-i\gamma_4\big)$.

We consider the fusion between $\gamma_2$ and $\gamma_3$. 
To analyze the outcome of fusing $\gamma_2$ and $\gamma_3$, we use the associative F-move to move from the $Z$-channel initialization to the $X$-channel by 
\begin{equation}
\begin{aligned}
    &|0_X\rangle=\frac{1}{\sqrt{2}}\left(|00\rangle+|11\rangle\right), \quad |1_X\rangle=\frac{1}{\sqrt{2}}\left(|00\rangle-|11\rangle\right), \\
    &|00\rangle=\frac{1}{\sqrt{2}}\left(|0_X\rangle+|1_X\rangle\right), \quad |11\rangle=\frac{1}{\sqrt{2}}\left(|0_X\rangle-|1_X\rangle\right)\label{eq:FmoveX}
\end{aligned}
\end{equation}
where $|0\rangle_X$ ($|1\rangle_X$) correspond to the arising of vacuum (fermion) in the $\rm{X}$-fusion channel, respectively.
See 
the Appendix for more information. 
To determine the charge arising from an arbitrary state during a $\mathcal{F}_X$ process, we first set $\langle n^{\rm{ref}}_i(T)\rangle=\langle 0_X|n_i|0_X\rangle$ for our reference state $\ket{0_X}$.
For an arbitrary single-qubit state, $|\psi(\theta,\phi)\rangle$, we again isolate the fusion result by considering the charge difference between the state and reference state, given by
\begin{equation}
    \begin{aligned}
     \Delta Q(T)&=\frac{-e}{2}\sum_i\langle \psi(\theta,\phi)(T)|n_i-\langle n^{\rm{ref}}_i\rangle |\psi(\theta,\phi)(T)\rangle\\
    &=-\frac{e}{2}\left(1+\rm{sin}(\theta)\rm{cos}(\phi)(\langle n^{\rm{ref}}_i\rangle-\langle 1_X|n_i|1_X\rangle)\right)\\
     &=-\frac{e}{2}(1-\langle \sigma_x\rangle)
     \end{aligned} \label{eq:Xcharge}
 \end{equation}
where we note that in this case, $\sum_i \left( \langle 1_X|n_i|1_X\rangle-\langle n^{\rm{ref}}_i\rangle\right)=1$, as the two eigenstates $\ket{0_X}$ and $\ket{1_X}$ differ by 1 fermionic charge after the fusion event. This assumption agrees with our numerical simulations, where we stress the fusion event is isolated from the additional MZMs in the system.
Unlike in the $Z$-channel, $\langle \sigma_x\rangle=i\langle \gamma_3\gamma_2\rangle$ is not diagonal in the basis $\{|00\rangle,|11\rangle\}$, leading to uncertainty in the final charge outcome. 
As such, if we set $|\psi(\theta,\phi)\rangle=|0\rangle_{\rm{logic}}$ or $|1\rangle_{\rm{logic}}$, $\Delta Q(\theta,\phi)(T)=-0.5e$ after the fusion event in both cases for the reference state $\ket{0_X}$. 
This has been denoted as a `nontrivial fusion' in previous works \cite{Zhou2022,Bai2024}, and, importantly, is indicative of the $\sigma \times \sigma= \mathbf{1}+\zeta$ fusion rule that allows for the non-Abelian statistical properties of the fusion space. 
For a generic state, $\ket{\psi(\theta,\phi)}$, the charge associated with the fusion result is given in Eq.\,\eqref{eq:Xcharge}.
We see this in Fig.\,\ref{fig:Fig2Kit}\,(b) and (c), where the charge density at the fusion site tends towards the same result in both cases, validating the non-Abelian statistical properties of the Ising anyon fusion algebra.

We now perform the same exercise as we did for the $Z$-Fusion and look to examine the charge quantization of the $\mathcal{F}_X$ fusion. 
This is done in Fig.~\ref{fig:Fig2Kit}\,(d) by again, conducting the fusion calculation for changing fusion time and delay coefficient.
We expect from Eq. \ref{eq:Xcharge}, that the states $|\mp\rangle_{\rm logic}=\frac{1}{\sqrt{2}}(|0\rangle_{\rm logic}\mp |1\rangle_{\rm logic})$ are the states diagonal in $\sigma_x$, and thus correspond to the fusion outcomes $\Delta Q=-e$ and $0$, respectively.
As mentioned before, we set $|+\rangle_{\rm logic}$ as reference state, as this corresponds to the vacuum fusion scenario. 
As such, we may isolate the charge of the fusion result 
and test the dependence of $\Delta Q_{|-\rangle}(T)$ on the speed of the process, as done in Fig.~\ref{fig:Fig2Kit}\,(d). 
While this yields a similar outcome to Fig.~\ref{fig:Fig1Kit}\,(d), with convergence in $\Delta Q_{|-\rangle}(T)$ occuring more quickly for smaller delay coefficients $\alpha$, we find one distinct difference: 
not all regimes converge to $\Delta Q_{|-\rangle}(T)=-e$. 
This cannot be explained simply due to diabatic errors, caused by mixing of the Majorana ground states with low-lying bulk-states, but rather reveals that the final fusion outcome materially depends on the dynamical procedure we fuse the two MZMs, which may lead to possible error in the final fusion outcome.
As shown in Fig.\,\ref{fig:Fig2Kit}\,(d), this emphasized by the fact that, surprisingly, convergence towards $-e$ occurs most clearly in the case $\alpha=0$, which corresponds to the case where we simultaneously ramp the chemical potential along each site during the fusion process, with $\Delta Q_{\ket{-}}(T)=-0.998e$ at $T=1140 \hbar/t$.
As we increase the delay coefficient from this point, the charge difference $\Delta Q_{\ket{-}}(T)$ deviates away from the expected outcome.

\begin{figure}[t!]
    \centering
\includegraphics{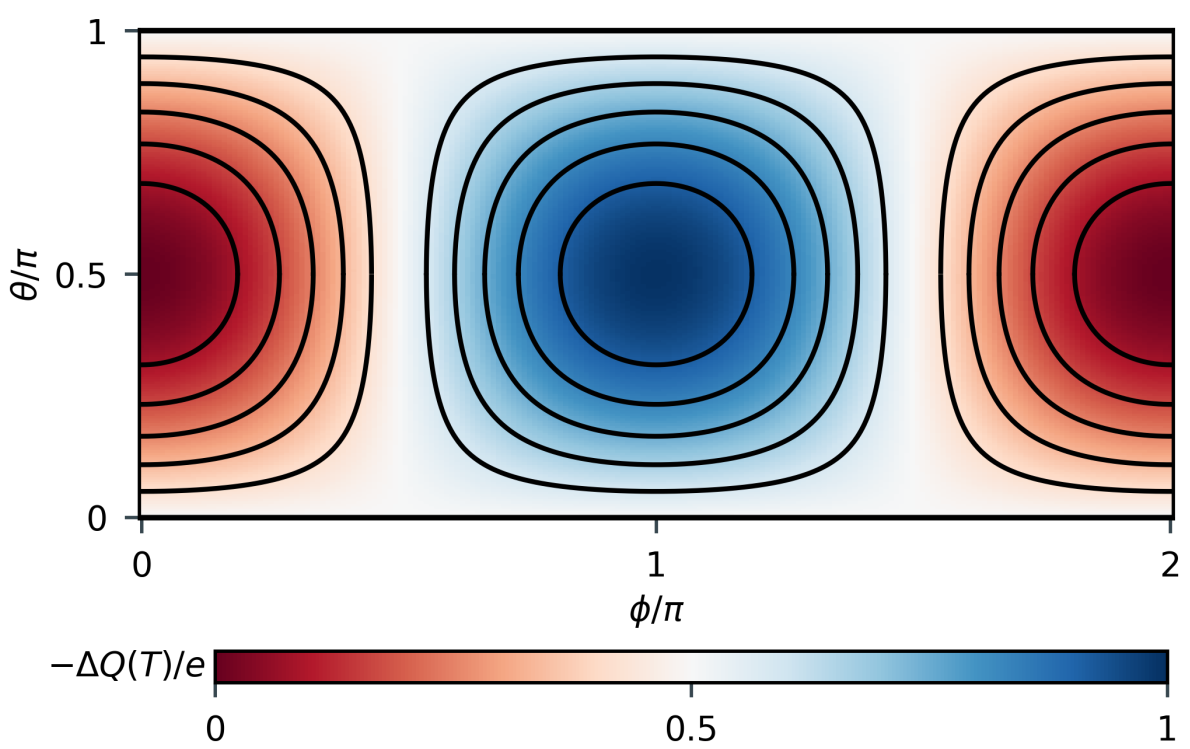}
    \caption{Phase readout through measurement of the charge difference in the X-channel. Charge difference, $\Delta Q(\theta,\phi)(T)$, between an arbitrary state $|\psi(\theta,\phi)\rangle_{\rm logic}$ and reference state $\ket{+}_{\rm logic}$ as a function of $\theta \in [0,\pi]$ and $\phi \in [0,2\pi]$.}
    \label{fig:Fig3Kit}
\end{figure}

We may now see clearly how we may probe the phase information through the X-channel. 
For an initial state $|\psi(\theta,\phi)\rangle$, we probe the $X$-channel charge readout for changing $\theta \in [0,\pi]$, $\phi \in [0,2\pi]$ in Fig.\,\ref{fig:Fig3Kit}, sweeping over all states along the Bloch-sphere. 
Here, we similarly calculate the charge difference $\Delta Q(\theta,\phi)(T)$ (with reference state $\ket{+}$).
For $\alpha=0$, we recover the maximal result $\Delta Q(\frac{\pi}{2},\pi)(T)=-0.998e$ at the $\langle\sigma_x\rangle=-1$ point ($\theta=\pi$, $\phi=0$). 
However we also gain constant charge along contours of constant $\langle \sigma_x \rangle$. 
This is demonstrated most clearly along $\phi/\pi=0.5,1.5$, where the charge converges to $-0.5e$, corresponding to $\langle\sigma_x\rangle=0$. 
Positively, this confirms the expected fusion result given in Eq.\,\eqref{eq:Xcharge}, confirming the X-channel fusion as a pathway towards a $\sigma_x$-readout in the adiabatic limit. 
While this confirms the analaytic result, we stress such a fusion via a ramp through the trivial regime presents clear limitations vs.\ the Z-channel case, with a near unity result in $\Delta Q_{|-\rangle}(T)$ much more difficult to achieve than the $\mathcal{F}_Z$ scenario.

%
\subsection{Braids} 
\begin{figure}[t!]
    \centering
\includegraphics{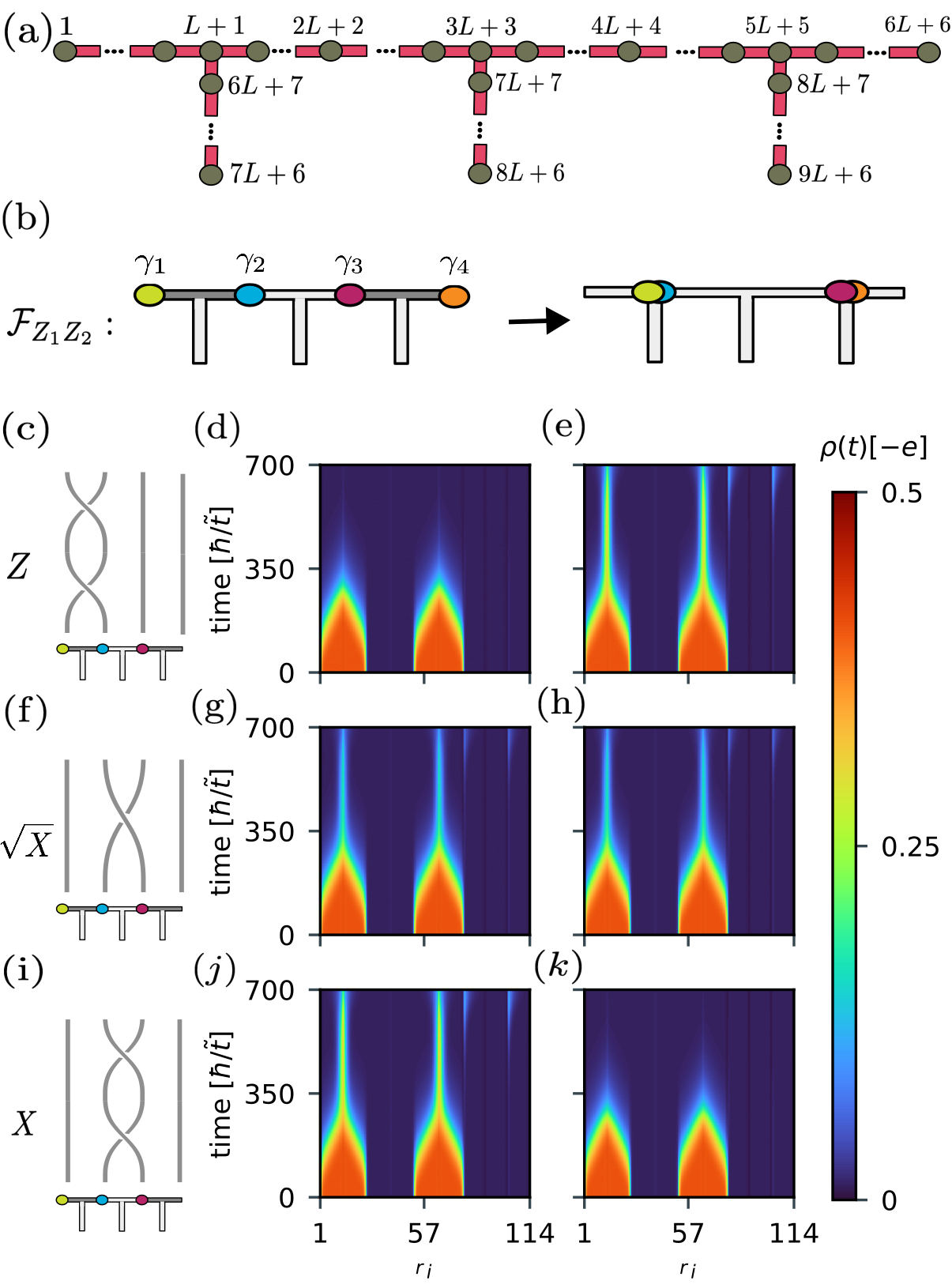}
    \caption{Fusion through the $Z$-channel after braids. (a) Definition of site labeling on the triple T-junction. (b) Graphic of $\mathcal{F}_{Z_1Z_2}$ fusion on a triple T-junction. (c, f, i) Braiding world lines for a $Z$, $\sqrt{X}$ and $X$ gate. (d,e), (g,h), (j,k)
    The local charge density, $\rho_i(t)$, along the triple T-junction, is plotted after a $Z$, $\sqrt{X}$, $X$ gate, respectively, where we fuse via the $\mathcal{F}_{Z_1Z_2}$ channel. 
    We initialize at $|00\rangle$ for (d, g, j) and $|11\rangle$ for (e, h, k). $(\tilde{t},\Delta,\mu_{\rm topo}, \mu_{\rm triv},L_{\rm leg})=(1,1,-0.75,-7.2,12a_0)$.
    Note that the time $t=0$ is directly after the braid.}
    \label{fig:Fig4Kit}
\end{figure}

%
We now test the effect of braiding on the fusion result through the $Z$-channel. Exchanging MZMs, as required for braiding, is not possible on a simple chain; instead, one uses a T-junction structure\,\cite{Alicea2011}, 
which allows to move one of the MZMs on an additional chain segment, so that the other MZM can swap its position. As a consequence, the world lines of the two MZMs ``cross'' over the wire network, performing the braid\,\cite{Mascot2023,Bedow2024,Hodge2024}.
This is given in Fig.\,\ref{fig:Fig4Kit}\,(a), where a triple T-junction, with leg lengths of size $L$, is illustrated. 
As given diagramatically in Fig.\,\ref{fig:Fig4Kit}\,(c, f, i), we conduct the $Z$, $\sqrt{X}$ and $X$ braid on a triple T-junction. 
As discussed in the Appendix, one of the fundamental advantages of Ising anyon-based models is that the spatial exchange of anyons leads to unitary gates on the ground-state manifold of the system. 
Such operations map the initialized qubit to a new quantum state, which is critical for the implementation of quantum computation. 
To show this, we perform a $\mathcal{F}_{Z_1Z_2}$ fusion on each Majorana bound state, calculating the expected charge density, $\rho_i(t)$, after each respective braid. 
We initialize the system with the physical states $|00\rangle$ (Fig.\,\ref{fig:Fig4Kit}\,(d, g, j)) and $|11\rangle$ (Fig.\,\ref{fig:Fig4Kit}\,(e, h, k)), which would lead to the arising of 0 and 2 fermions respectively through the $\mathcal{F}_{Z_1Z_2}$ channel, respectively.
In the case of a $Z$-gate (Fig.\,\ref{fig:Fig4Kit}\,(c-e)), which will only induce a phase difference between the two states, we expect no transfer in the charge associated with each Majorana bound states after the braid.
As given in Fig.\,\ref{fig:Fig4Kit}\,(d), we measure no charge for the $|00\rangle$ state after fusion, but see a clear signal for the $|11\rangle$ state, as demonstrated in Fig.\,\ref{fig:Fig4Kit}\,(e), the expected result in the $Z$-basis.
This is in contrast to the $X$-gate (Fig.~\ref{fig:Fig4Kit}\,(i-k)), where we see a swap in the charge distribution, with the $|00\rangle$ ($|11\rangle$) state now increasing (decreasing) the charge signal after the fusion result. 
This agrees with the expected Ising anyon braid statistics (see Appendix for details), where the $X$-procedure corresponds to a direct charge transfer between the $|00\rangle$ and $|11\rangle$ state. Further, the result remains well quantized, much like in the $Z$-fusion case, with $\Delta Q(\pi,0)(T)$=$-1.991e$ for the $Z$-braid (Fig.\,\ref{fig:Fig4Kit}\,(d, e)) and $X$-braid (Fig.\,\ref{fig:Fig4Kit}\,(j, k)), respectively. 
Importantly, we also see an equal charge distribution after the $\sqrt{X}$-gate (Fig.\,\ref{fig:Fig4Kit}\,(f-g)), with the final charge difference between the $|00\rangle$, $|11\rangle$ initial states, $\Delta Q(T)=-5.40\times 10^{-3}e$, confirming that both charge measurements are almost identical. 
Much like the fusion results through the $X$-channel result, we gain an equal superposition between the two fusion states $\ket{00}$ and $\ket{11}$, giving equality in the charge density calculated through the $Z$-channel.
Previous works have calculated transition probabilities, $p_{ij}=|\langle i|j(t)\rangle|^2$, to numerically demonstrate the success of a dynamic braiding procedure \cite{Sekania2017,Sanno2021,Bedow2024,Mascot2023}. 
Here, we demonstrate the relevant braid statistics using a charge readout, not only confirming the relevant unitary transformations, i.e., $\sqrt{Z}$, $\sqrt{X}$ and $H$ gates, on the fusion space (see the Appendix for further details on the gate definitions used in this work) but demonstrating how these braid statistics can be interpreted in terms of a charge measurement, a measurable quantity in the context of a physical topological quantum computer. 
\begin{figure}[t!]
    \centering
    \includegraphics{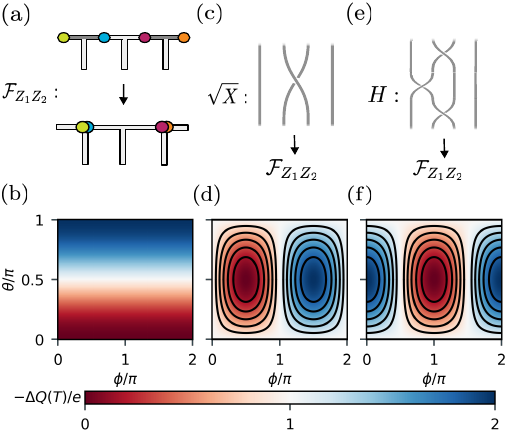}
    \caption{Bloch sphere readout through the $Z$ channel. (a, c, e) Routine required for $\langle \sigma_z \rangle$, $\langle \sigma_y \rangle$, $\langle \sigma_x\rangle$ readout, respectively. 
    (b, d, f) The final charge difference $\Delta Q(\theta,\phi)(T)$ after the $\mathcal{F}_{Z_1Z_2}$ routine for the $\langle \sigma_z \rangle$, $\langle \sigma_y \rangle$, $\langle \sigma_x\rangle$ readout. We plot the charge difference, taken between an arbitrary state $|\psi(\theta,\phi)\rangle_{\rm logic}$ and the reference states 
     $|0\rangle_{\rm logic}$, $\frac{1}{\sqrt{2}}(|0\rangle_{\rm logic}+i|1\rangle_{\rm logic})$, $\frac{1}{\sqrt{2}}(|0\rangle_{\rm logic}-|1\rangle_{\rm logic})$, respectively, as a function of $\theta$ and $\phi$. The parameters used for these simulations are $(\tilde{t},\Delta,\mu_{\rm topo}, \mu_{\rm triv},L_{\rm leg})=(1,1,-0.75,-7.2,12a_0)$.}
    \label{fig:Fig5Kit}
\end{figure}
With the charge transfer matching the Ising anyon braid statistics, we now look to see how we can utilize the braiding statistics on the fusion space for the purpose of a full state readout.
Utilizing Eq. \eqref{eq:Braid} on the $\{|00\rangle,|11\rangle\}$ subspace, for an arbitrary single qubit state $|\psi\rangle=\textrm{cos}(\frac{\theta}{2})|00\rangle +e^{i\phi}\textrm{sin}(\frac{\theta}{2})|11\rangle$, with $\theta\in [0,\pi]$, $\phi \in [0,2\pi]$, we find that the expected charge associated with the fusion outcome, $\Delta Q(\theta,\phi)$, is given by
\begin{align}
    &\mathcal{F}_{Z_1Z_2}: \quad \quad \quad \; \; \;  \Delta Q(\theta,\phi)(T)=-e(1-\langle \sigma_z\rangle), \label{eq:readoutZ}\\[5pt]
    &\mathcal{F}_{Z_1Z_2}H: \quad \quad \quad  \Delta Q(\theta,\phi)(T)=-e(1+\langle \sigma_x\rangle), \label{eq:readoutX}\\ 
    &\mathcal{F}_{Z_1Z_2}\sqrt{X}: \quad \quad \: \Delta Q(\theta,\phi)(T)=-e(1-\langle \sigma_y\rangle).
    \label{eq:readoutY}
\end{align}
Here $\langle \sigma_x\rangle=\textrm{cos}(\theta)\textrm{sin}(\phi)$, $\langle \sigma_y\rangle=\textrm{sin}(\theta)\textrm{sin}(\phi)$ and $\langle\sigma_z\rangle=\rm{cos}(\theta)$.
This provides access to not only the population of the $|11\rangle$ state, given by the $\sigma_z$ projection, but also the phase information $\textrm{tan}(\phi)=\langle \sigma_y\rangle/\langle\sigma_x\rangle$. 

We again isolate the charge associated with the fusion by finding the charge difference between an arbitrary state $|\psi(\theta,\phi)\rangle_{\rm logic}$ and a reference state which corresponds to the vacuum fusion result. 
For the $\mathcal{F}_{Z_1Z_2}$, $\mathcal{F}_{Z_1Z_2}H$ and $\mathcal{F}_{Z_1Z_2}\sqrt{X}$ routines, respectively, this zero-charge reference state will be given by $|\psi\rangle=|00\rangle$, $\frac{1}{\sqrt{2}}\left(|00\rangle-|11\rangle\right)$, $\frac{1}{\sqrt{2}}\left(|00\rangle+i|11\rangle\right)$, aligning with the positive $z$, and negative $x$, $y$ axis, respectively, on the Bloch sphere. 

We begin with the $\mathcal{F}_{Z_1Z_2}$ operation, given in Fig.\,\ref{fig:Fig5Kit}\,(a) and (b). 
Here, we clearly see a transition from $\Delta Q(T)=0$ to $-2e$ as $\theta$ transitions from $0$ to $\pi$, with $\Delta Q(\pi,0)(T)=-1.991e$, thus giving near quantization in the calculated charge difference. 
Further, there is no change in the charge measurement as we change $\phi$, suggesting a successful $\langle \sigma_z\rangle$ measurement.

We extend this to a phase measurement, as given by Fig.\,\ref{fig:Fig5Kit}\,(c)-(f). 
For the $\langle\sigma_y\rangle$ calculation in Fig.\,\ref{fig:Fig5Kit}\,(d), we see clear extrema forming at $\theta=\frac{\pi}{2}$, $\phi=\frac{\pi}{2},\frac{3\pi}{2}$, with $\Delta Q(T)=0,-1.991e$ respectively. 
These agree with the analytical Eqs.\,\eqref{eq:readoutX} and \eqref{eq:readoutY}, where each point will lead to a charge readout of $0,2$, respectively. 
We repeat the procedure for the $\sigma_x$ case (Fig.\,\ref{fig:Fig5Kit}\,(f)), where here we see extrema forming instead at $\theta=\frac{\pi}{2}$, $\phi=0,\pi,2\pi$, also with charge difference $\Delta Q(T)=-1.990e$ at the maximum. 
These not only match well with the expected result, but we also see a distinct advantage in a phase readout through the $Z$-channel instead of the $X$-channel. 
As discussed in Sec.\,\ref{sec:Z-fusion}, we find that fusion through the $X$-channel, while also another feasible method to predict the phase of $|\psi\rangle_{\rm logic}$, we required both large times $T$ and in the calculated range $\alpha=0$ in order to get a near quantized result. 
Thus, we clearly see a comparative advantage in fusing through the $Z$-channel when looking for a full state readout. 

%
\subsection{Charge Localization on a Quantum Dot} 
%

While previous results have clearly demonstrated charge readouts near quantization, both in the $Z$ and $X$ basis, we run into a fundamental problem.
How do we isolate the charge for the purpose of a readout mechanism?
Charge readout schemes such as a charge capacitance readout\,\cite{Zhou2022,Liu2023,Smith2020}, charge pumping schemes\,\cite{Souto2022}, along with weak measurement proposals\,\cite{Steiner2020,Bai2024,Wang2024} all require the fusion result to remain local, thus allowing for capture of the fusion result before the charge is able to roam freely in the bulk. 

To this end, we consider the coupling of Kitaev chains to an external quantum dot (QD);
more accurately, one could call it simply a defect site or point defect, but in line with previous work\,\cite{Steiner2020,Sau2024,Boross2024} we will refer to it as QD (we note that a physical QD would correspond to an extended size, i.e. at least a few extra sites and not just a single site). 
 Here, in a full dynamic simulation, we aim to test the following: can we use such an infrastructure to isolate the fusion result on the QD, and will we recover the full charge $\langle Q(T)\rangle=-ne$ on the QD.

\begin{figure}[t!]
    \centering
    \includegraphics{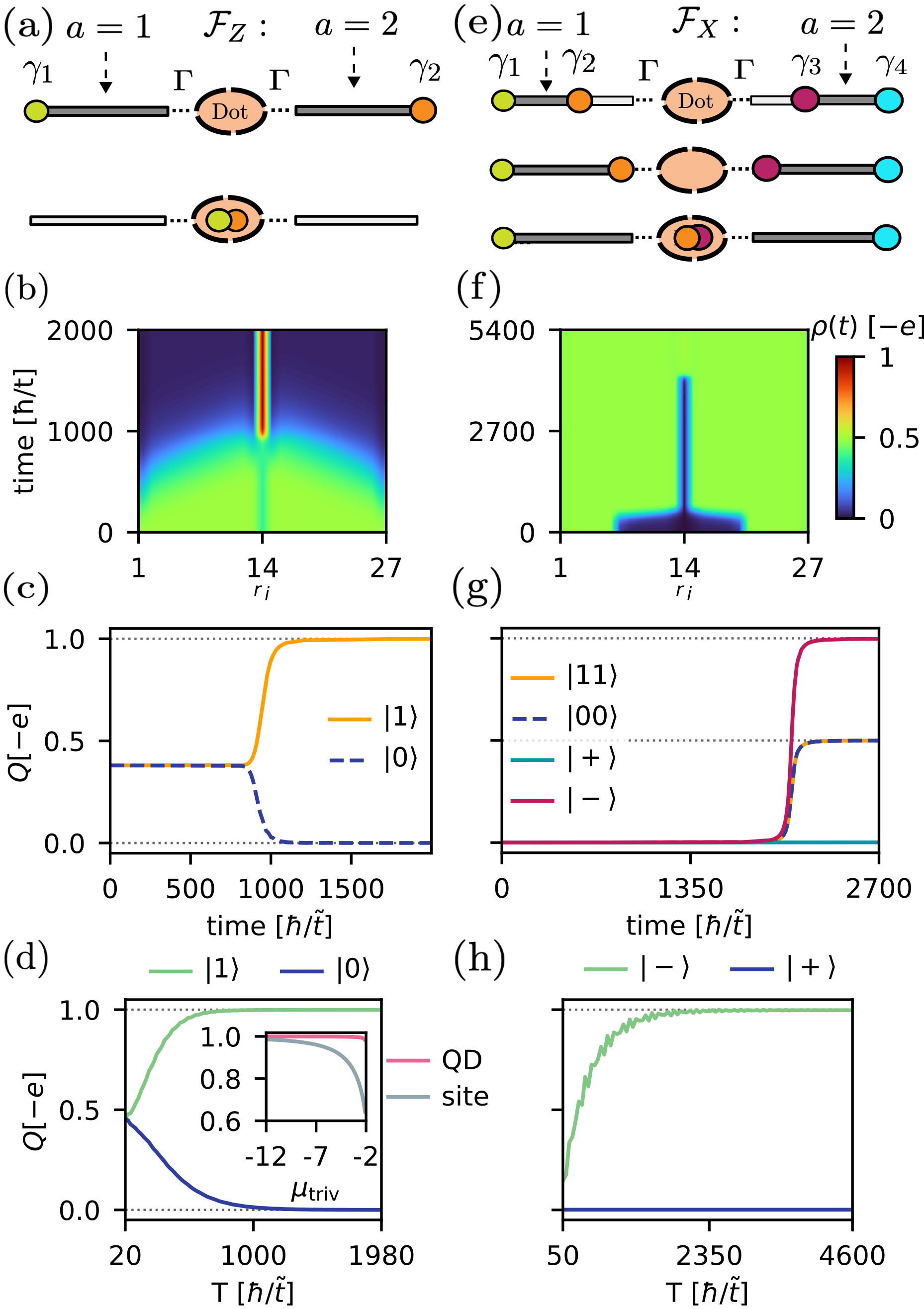}
    \caption{Fusion on a quantum dot. (a-d) Results for the $\mathcal{F}_Z$ fusion process. 
    (e-h) Results for the $\mathcal{F}_X$ fusion process.  (a), (e) Graphic of each fusion process involving the QD.
    Grey (white) denotes the chain in the topological (trivial) regime.
    (b), (f) Local charge density $\rho_i(t)$ over the fusion process for two Kitaev chains of length $L=13 a_0$. 
    (c), (g) Charge $Q_{\rm{QD}}(T)$ accumulated on the QD as a function of time $T$ over the fusion process. 
    (d), (h) Final charge accumulated on the QD as a function of the total time of the process, with the inset in (d) providing a comparison between the charge localized on a single topological site and a QD for $T=4000\frac{\hbar}{\tilde{t}}$ for initial state $|1\rangle_{\rm logic}$. 
    For (b-d), $(\mu_{\rm{topo}},\mu_{\rm{triv}},\mu_{\rm{QD}},\Delta,\tilde{t},\Gamma)=(-0.05,-4.5,1,1,0.1)$. For (f-h), $(\mu_{\rm{topo}},\mu_{\rm{triv}},\mu_{\rm{QD}},\Delta,\tilde{t},\Gamma)=(-0.1,-4.5,1,1,0.05)$ with the final $\mu_{\rm QD}=1\tilde{t}$ in the $\mathcal{F}_X$ scenario.}
    \label{fig:Fig6Kit}
\end{figure}

The Hamiltonian now takes the form:
\begin{equation}
\begin{aligned}
    \mathcal{H}_{\rm QD}(t) =& \, \mathcal{H}_{\rm{Kit}}(t)-\sum_{a\in \{1,2\}}\Gamma \left( c^{\dag}_{L,a}c_{\rm{QD}}+c^{\dag}_{\rm{QD}}c_{L,a} \right) \\
    &-\mu_{\rm{QD}}(t)c^{\dag}_{\rm{QD}}c_{\rm{QD}}, \label{eq:QD}
\end{aligned}
\end{equation}
 where the $\Gamma$ corresponds to an electrostatic coupling to the ends of two Kitaev wires, with the QD tuned with a local chemical potential $\mu_{\rm{QD}}$, and $a=1, 2$ labels the two Kitaev chains. 
A graphic of this is provided in Fig.\,\ref{fig:Fig6Kit}\,(a) and (e), with the QD itself acting as a defect which will serve to capture the charge when the MZMs are neighboring it.

We begin with a fusion in the $Z$-basis. 
On a system containing a single pair of MZMs, we ramp the MZMs onto the QD, as indicated in Fig.\,\ref{fig:Fig6Kit}\,(a), while setting $\mu_{\rm{QD}}=\mu_{\rm topo}=-0.1\tilde{t}$ throughout the time-evolution. 
As the chemical potential of the chains are ramped from $\mu_{\rm topo}$ to $\mu_{\rm triv}$, the MZMs are displaced off of the now trivial Kitaev wires, and onto the QD. 
For $\Gamma=0$, this system would correspond to two distinct Kitaev chains, with MZMs arising on all four boundaries. 
However, as $\Gamma$ increases, the MZMs neighbouring the QD hybridize, with the associated Majorana bound state splitting from the MZM subspace, forming a higher energy bound state. 
As can be seen in Fig.\,\ref{fig:Fig6Kit}\,(b), this leads to a distinct charge signal for the filled $|1\rangle$ state, on the QD itself, located at $r_i=13 a_0$, thus demonstrating the localization of the charge result. 

We now aim to confirm the convergence of the charge readout with the expected outcome. 
In all cases where we calculate the final charge, we will only consider the charge on the QD, given by $Q_{\rm{QD}}(T)$.
This is given in Fig.~\ref{fig:Fig6Kit} (c), where we see not only a clear convergence of the charge to $Q_{\rm{QD}}(T)=-e$, clearly indicating the arising of a massive fermion, $\zeta$, locally on the QD, but also the convergence of the $|0\rangle$ state to $Q_{\rm{QD}}(T)=0$, confirming both quantized results in the diagonal $\sigma_z$ basis. 
We further see that, as in Fig.~\ref{fig:Fig1Kit} (d), Fig.~\ref{fig:Fig6Kit} (d) demonstrates a clear tending towards the $Q_{\rm{QD}}(T)=-e$ $(0)$ as we approach the adiabatic limit for the $|1\rangle$ ($|0\rangle$) state. This clearly demonstrates that, under this protocol, we are able to recover the fusion result in the adiabatic limit.
Critically, the fusion result will remain centred on the QD, essential for the purpose of a readout.   
We further see that, while an additional subgap state on the boundary of the topological regimes next to the QD, in the adiabatic limit, this will not lead to a loss of quantum information due to quasiparticle poisoning, preserving the logical subspace.

Further, in the inset of Fig.\,\ref{fig:Fig6Kit}\,(d), we compare this localization procedure with another scenario, where we replace the QD with a single topological site impurity, where the chemical potential of the site is set to $\mu_{\rm{topo}}$.
By similarly ramping the MZMs towards this site, this traps the MZMs at a single defect in the chain, whereby the potential gradient between $\mu_{\rm{topo}}$ and $\mu_{\rm{triv}}$ similarly traps the MZMs on a single site, thus localizing the fusion result.
However, as shown in the inset, by calculating the charge on the fusion site for both processes, we reveal a distinct advantage in the QD localization scheme.
In the adiabatic limit (fusion time $T=4000\hbar/\tilde{t}$), this localization procedure clearly converges towards analytical expectations as $\mu_{\rm{triv}} \to -12\tilde{t}$, with the large potential gradient in this case effectively isolating the MZM wavefunctions on the topological site. As $\mu_{\rm{triv}}$ decreases, the measured charge on the fusion site tends away from $-e$.
This is in clear contrast to the QD, whereby we see a clear convergence of the charge to $-e$ over the entire topological regime, with $Q_{\rm{QD}}(T)/(-e)>0.99$ in the regime where $\mu_{\rm{triv}}\leq-2.3\tilde{t}$. 

We can extend our considerations to the $\mathcal{F}_X$ fusion scenario. 
In this case, we need to adjust the fusion protocol, as we instead manipulate the position of the MZMs by ramping from $\mu_{\rm triv}$ to $\mu_{\rm topo}$, as shown in Fig.~\ref{fig:Fig6Kit} (e). 
This no longer guarantees the displacement of the MZMs, as each Kitaev chain will individually remain in the topological regime. 
Instead, we attempt to capture this charge by additionally driving the potential barrier of the QD, such that it becomes energetically favourable for the MZMs to occupy the QD as compared to remaining on the chains, thus allowing for fusion to take place on the QD.

We do this by initializing $\mu_{\rm QD}(0)=\mu_{\rm triv}$. 
In this case which initializes the energy of the dot well within the bulk states of the spectra, above the Majorana subspace. 
As displayed graphically in Fig.\,\ref{fig:Fig6Kit}\,(e), after we ramp the Kitaev chains from trivial to topological, $\gamma_2$ and $\gamma_3$ remain on the boundaries of the chains.

In order to isolate them onto the dot, we then ramp $\mu_{\rm QD}$ from $\mu_{\rm triv}$ to $\mu^f_{\rm QD}=\tilde{t}$. 
We do this by using the same polynomial interpolation given in Eq.\,\eqref{eq:ramp}, in this case setting $\mu_{\rm topo}\to \mu^f_{\rm QD}$.
This procedure brings the energy on the dot below the Majorana subspace, leading to the lowest energy configuration to correspond to the filling of the QD, isolating the MZMs on the external structure. 
We confirm this in Fig.\,\ref{fig:Fig6Kit}\,(f) and (g), where we see the charge of the QD for the initial $|11\rangle$ state to converge to $Q_{\rm{QD}}(T)=-0.5e$ over the fusion process. 
We confirm this in (g), where both the $|00\rangle$ and $|11\rangle$ converge to $Q=-0.5e$, with the states diagonal in $\sigma_x$, $|\pm\rangle_{\rm logic}=\frac{1}{\sqrt{2}}(|00\rangle\pm |11\rangle)$, converging towards $Q_{\rm{QD}}(T)=0$ for $\ket{+}_{\rm logic}$ and towards $Q_{\rm{QD}}(T)=-e$ for $\ket{-}_{\rm logic}$. 
This is in agreement with both analytical expectation, and the results discussed in Fig.~\ref{fig:Fig3Kit}. 
We again perform an adiabaticity test, as given in Fig.~\ref{fig:Fig6Kit} (h), where we see convergence of the charge for the $|-\rangle$ state towards $-e$ in the adiabatic limit, in agreement with results in the previous section. 

As such, for both fusion processes, we can enact a fusion protocol which isolates the charge on the QD, and approaches the analytical result in the adiabatic limit.  


%
\section{Fusion on Magnet-Superconductor Hybrid Systems}\label{sec:MSH}

\begin{figure}[t!]
  \centering
  \includegraphics[width=\columnwidth]{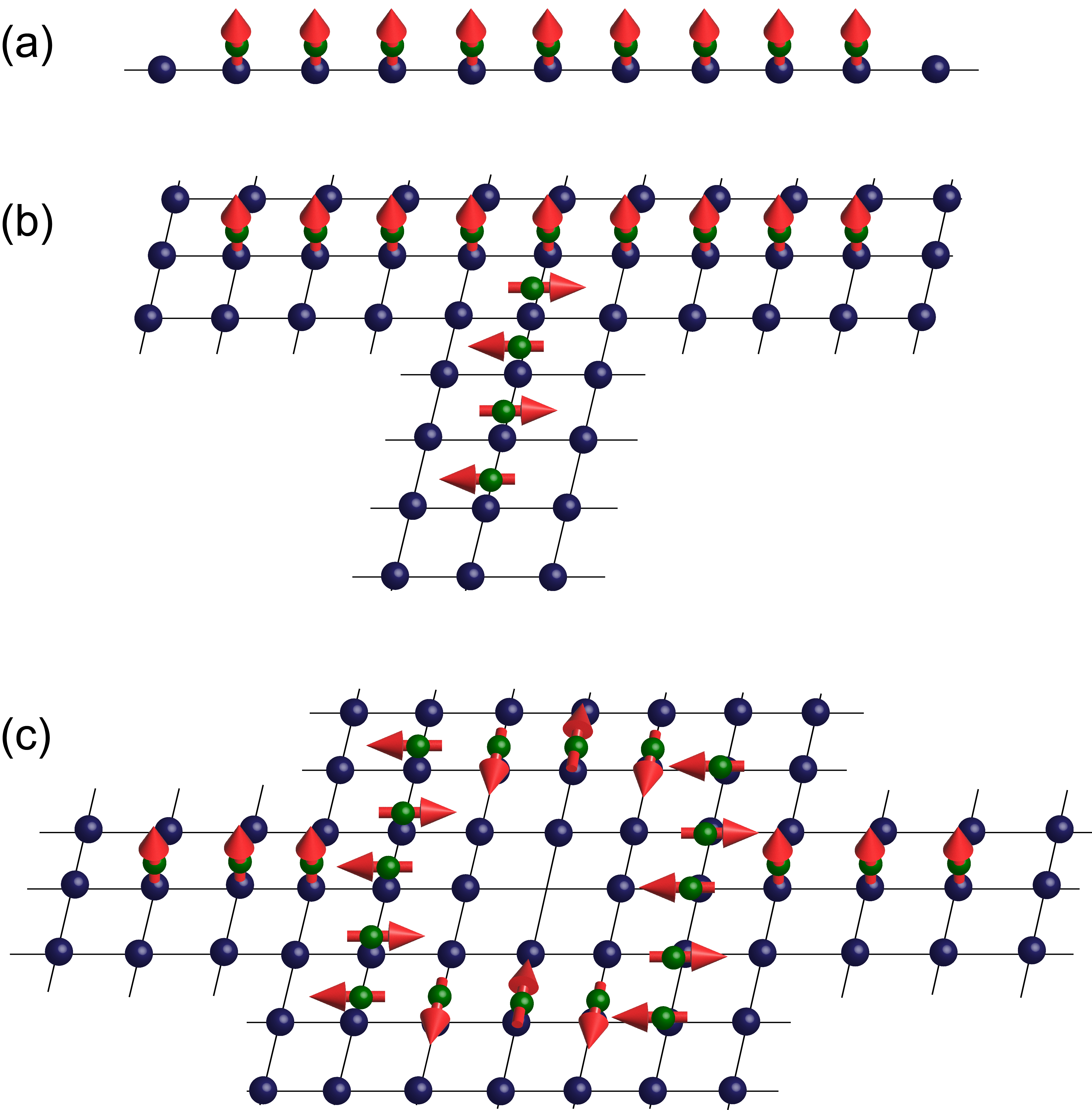}
  \caption{
  Schematic of MSH systems with magnetic adatoms placed on the two-dimensional, square lattice surface of an $s$-wave supercondutor:
  (a) a one-dimensional chain,  
  (b) a T-structure, and 
  (c) a square-and-arms structure.}
  \label{fig:Fig7}
\end{figure}

Strong evidence for the existence of topological superconductivity and Majorana zero modes have been reported in 1D\,\cite{Nadj-Perge_2014, Ruby2015, Pawlak2016, kim_toward_2018, rachel-25pr1}
and 2D\,\cite{menard_two-dimensional_2017, PalacioMorales2019, kezilebieke_topological_2020, bazarnik_antiferromagnetism-driven_2023, soldini_two-dimensional_2023, bruning_non-collinear_2024, loconte-25nc}
MSH systems, in which magnetic adatoms are placed on the surface of an $s$-wave superconductor.
It was recently theoretically demonstrated that MZMs in 1D MSH networks, such as the ones shown in Fig.\,\ref{fig:Fig7}, can be braided and utilized for the creation of quantum algorithms by manipulating the local magnetic structure\,\cite{Bedow2024,Bedow2025}. In particular, by rotating the magnetic moments between an out-of-plane ferromagnetic (FM), and in-plane antiferromagnetic (AFM) alignment, an MSH network can be tuned between topological (for the FM alignment) and trivial (for the AFM alignment) phases. 
This proposal was motivated by recent electron spin resonance -- scanning tunneling microscopy (ESR-STM) experiments that demonstrate that the spins of individual target magnetic adatoms can be flipped depending on the spin orientation of a control adatom \cite{Yang2019, Wang2023, Phark2023}. 
Moreover, it was shown that the braiding process can be entirely visualized using the energy-, time-, and spatially resolved non-equilibrium density of states, $N_\mathrm{neq}$, which was shown to be proportional to the time-dependent differential conductance measured in scanning tunneling spectroscopy experiments\,\cite{Bedow2022}. 

Below, we visualize the braiding and fusion processes using $N_\mathrm{neq}$ and demonstrate that the read-out of the final qubit state is possible using the time-dependent excess charge density after fusion.  
In that sense,  we will look to probe how the physics of MZM fusion transfers from the spinless $p$-wave Kitaev system to the experimentally viable spinful MSH system. 
Here, additional considerations such as wavefunction decay into the two-dimensional substrate, Rashba spin-orbit coupling, and the retention of a topological gap in the presence of an $s$-wave coupling \emph{must} be taken into account. 
However, as we show, since the low-energy physics of both models are essentially equivalent, in the adiabatic limit, the fundamental physics of MZM fusion will hold between the two systems.


%
\subsection{Theoretical Model}
%
To investigate the braiding and fusion of MZMs in MSH systems, we start from the Hamiltonian 
\begin{align}
    \mathcal{H}_{\rm MSH} = &-\tilde{t} \sum_{\mathbf{r}, \mathbf{r'},\sigma}
    c^{\dagger}_{\mathbf{r},\sigma}c_{\mathbf{r'},\sigma}
    -\mu\sum_{\mathbf{r},\sigma}c^{\dagger}_{\mathbf{r},\sigma}c_{\mathbf{r},\sigma}\nonumber \\
   &+\mathrm{i}\alpha\sum_{\mathbf{r},\boldsymbol{\delta},\sigma,\sigma'}
    c^{\dagger}_{\mathbf{r},\sigma}
    \left(
    \boldsymbol{\delta}\times\boldsymbol{\sigma}
    \right)^{z}_{\sigma,\sigma'} c_{\mathbf{r}+\boldsymbol{\delta},\sigma'} \nonumber \\
    &+\Delta\sum_{\mathbf{r}}
    \left(
    c^{\dagger}_{\mathbf{r},\uparrow}c^{\dagger}_{\mathbf{r},\downarrow}
    +   c_{\mathbf{r},\downarrow}c_{\mathbf{r},\uparrow}
    \right) \nonumber \\
    &+\sum_{\mathbf{R},\sigma,\sigma'}
    c^{\dagger}_{\mathbf{R},\sigma}
    \left[
    J_{\mathbf{R}}\mathbf{S}_{\mathbf{R}}(t)
    \cdot \boldsymbol{\sigma}
    \right]_{\sigma,\sigma'}
    c_{\mathbf{R},\sigma'}.
    \label{eq:H_MSH}
\end{align}

\begin{figure*}[]
  \centering
  \includegraphics[width=0.9\textwidth]{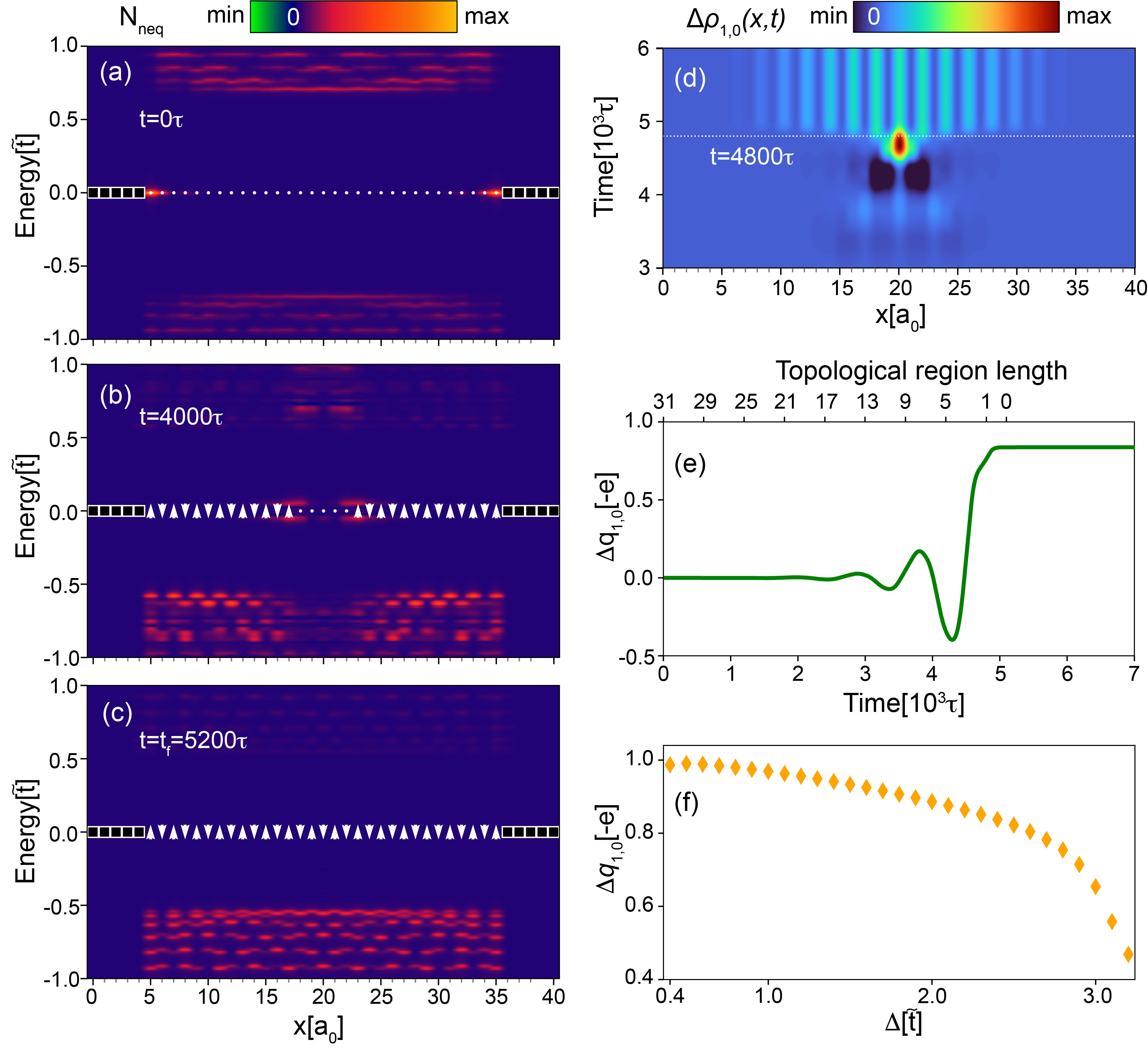}
  \caption{Energy-resolved non-equilibrium local density of states $N_\mathrm{neq}$ for different times during the fusion process: (a) at the beginning at  $t= 0 \tau$, (b) when the MZMs are only separated by a few out-of-plane spins at at $t=4000 \tau$, and
  (c) when the MZMs are completely fused at $t=5200 \tau =t_f$. The spin orientations are represented by white arrows, with white dots indicating an out-of-plane orientation. Black rectangles denote regions where the substrate is not covered with magnetic adatoms. The length of the substrate is 41 sites, with 31 magnetic adatoms placed at the center of the chain.
  (d)  Charge density difference $\Delta \rho_{1,0}(x,t)$ as a function of position along the chain and time. 
  (e) Total charge difference $\Delta q_{1,0}(t)$ as a function of time (bottom axis) and length of topological region (i.e., number of out-of-plane spins).
  (f) $\Delta q_{1,0}(t_f)$ after the completion of the fusion process as a function of the superconducting order parameter.
  Parameters are $(\mu, \alpha, \Delta, J) = (-4.0, 0.9, 2.4, 5.2)\tilde{t}$ with the rotation time $t_R = 600 \tau$ for one spin and the delay time $\Delta t_R = 300 \tau$ 
  until the rotation of the next spin. 
}
  \label{fig:Fig8}
\end{figure*}
Here, $c^{\dagger}_{\mathbf{r},\sigma}$ creates an electron at site $\mathbf{r}$ with spin $\sigma$. $\tilde{t}$ is the nearest-neighbor hopping amplitude in a two-dimensional square lattice, $\mu$ is the chemical potential, $\alpha$ is the Rashba spin-orbit coupling between nearest-neighbor sites connected by the vector $\boldsymbol{\delta}$, $\Delta$ is the $s$-wave superconducting order parameter. $J$ is the magnetic exchange coupling between the magnetic adatom with spin $S_\mathbf{R}(t)$ at site $\mathbf{R}$ and time $t$ and the conduction electrons, and $\boldsymbol{\sigma} = (\sigma_x, \sigma_y, \sigma_z)^T$ is the vector of Pauli matrices. 
Further, in this work, $\alpha$ and $\Delta$ have been chosen to maximize the topological gap and subsequently, minimize the localization length of the MZM wavefunctions.
This is done in order to conduct our numerical simulations on computationally viable system sizes without the detrimental effects of MZM hybridization while conducting quantum gates \cite{Hodge2024}. 
On larger system sizes, where more physical parameter choices may be chosen, we expect the conclusions made in this section to be qualitatively equivalent as the MZMs remain protected by the gap in both cases.
Since Kondo screening is suppressed by the hard superconducting gap in the $s$-wave superconductors \cite{Balatsky2006, Heinrich2018}, we assume the spins to be classical in nature with $S_\mathbf{R}(t)$ indicating the spin direction. 
A detailed discussion of this model, and its potential experimental realization for quantum gate operations is given in Ref.\,\cite{Bedow2024}.

In the MSH systems considered below, an out-of plane FM alignment of the magnetic adatoms' spins induces a topological phase, while an in-plane AFM alignment yields a trivial, but gapped phase. Since the MZMs are localized at the intersection of (or domain wall between) the topological and trivial regions of the MSH system, the MZMs can be moved by locally changing the spin configuration between these two alignments. 
Crucially, this rotation of individual magnetic moments has been realized experimentally via STM-ESR techniques \cite{Yang2019,Wang2023,Wang2024}, providing an exciting avenue to dynamically adjust the topological domain and thus, move MZMs through the MSH system. This provides the mechanism necessary to enable braiding and fusion in our simulations.
While the effects of any exchange interaction between neighboring adatoms is beyond the scope of this work, this interaction is diminished by utilizing more dilute configurations of magnetic adatoms, where topological phases have been shown to be retained \cite{Rontnyen2015,Rontnyen2016} and recent experimental work demonstrating the retention of boundary modes \cite{jang2025}.
How the physics of dynamic MZM braiding and fusion is affected by such a choice provides an interesting avenue towards future work. 
This rotation of magnetic moments is characterized by a rotation time $t_R$ to rotate a single moment by $\pi/2$ between in- and out-of-plane alignment and a delay time $\Delta t_R$ before the next spin begins its respective rotation. 
To ensure the adiabaticity of the entire process, we choose a rotation time $t_R \gg \hbar/\Delta_t$ where $\Delta_t$ is the topological gap in the system. 
Below all times are given in units of $\tau = \hbar/\tilde{t}$ such that for typical values of $\tilde{t}$ of a few hundred meV, $\tau$ is of the order of a few femtoseconds.

A fusion process is achieved by moving the domain walls, and hence the MZMs, towards each other. This process can be visualized by using the non-equilibrium local density of states defined via
\begin{align}
    N_\mathrm{neq}(\mathbf{r},\sigma, t, \omega) = 
    -\frac{1}{\pi}\mathrm{Im} \; G^{r}(\mathbf{r},\mathbf{r}, \sigma, t,\omega) \; ,
    \label{eq:N_neq}
\end{align}
which, as previously shown, is proportional to the time-dependent differential conductance\,\cite{Bedow2022}.
Here, $G^{r}(\mathbf{r},\mathbf{r}, \sigma, t,\omega)$ is a diagonal element of the time- and energy-dependent retarded Green's function matrix ${\hat G}^{r}(t,\omega)$ which can be obtained from the solution of the differential equation
\begin{align}
    \left[ \mathrm{i} \frac{d}{dt} +  \omega  + \mathrm{i} \Gamma - {\hat H}_{\rm MSH}(t)\right] {\hat G}^r \left(t, \omega \right) = {\hat 1} \; ,
    \label{eq:diff_Nneq}
\end{align}
where ${\hat H}_{\rm MSH}$ is the matrix representation of the Hamiltonian from Eq.~(\ref{eq:H_MSH}).

Finally, to read out the final state of the qubit, we compute the time-dependent local charge density of a state $\ket{i}$ given by  
\begin{equation}\label{eq:charge_dense}
 \rho_{\ket{i}}({\bf r},t) = - e \sum_{\sigma=\uparrow,\downarrow} \braket{i(t)|c^\dagger_{{\bf r}, \sigma} c_{{\bf r}, \sigma}|i(t)} \; ,
\end{equation}
where $\ket{i(t)}$ is the time-dependent many-body state in Fock space constructed according to Eq.\,\eqref{eq:Time-evolution}. Moreover, we define the charge density difference between two states $\ket{i}$ and $\ket{j}$ via
\begin{equation}
    \Delta \rho_{i,j}({\bf r}, t) = \rho_{\ket{i}}({\bf r},t) - \rho_{\ket{j}}({\bf r},t) \ ,
    \label{eq:rho_diff}
\end{equation}
and the corresponding time-dependent total charge difference via
\begin{equation}
    \Delta q_{i,j}(t) = \sum_{\bf r} \Delta \rho_{i,j}({\bf r}, t) \ ,
\end{equation}
also referred to as the excess charge.

%
\subsection{One-dimensional MSH structure}
%

We begin by considering the fusion of two MZMs in a one-dimensional MSH system consisting of a chain of magnetic adatoms, placed on top of a one-dimensional superconducting substrate. In the initial configuration, the spins are aligned ferromagnetically out-of-plane with the zero-energy MZM localized at the chain's ends  as shown in the energy-and spatially resolved $N_\mathrm{neq}$ in Fig.~\ref{fig:Fig8}(a). Here, the spin orientations are represented by white arrows, with white dots describing an out-of-plane alignment. Note that the black squares to the left and right of the chain represent sites that are not covered by magnetic adatoms. To fuse the two MZMs, we rotate the magnetic moments into the plane starting from the chain's ends -- yielding an AFM alignment -- and progressing towards the center of the chain. Here, we define the length of the topological region at any time as the number of spins oriented FM out-of-plane. 

As the MZMs approach each other, they begin to hybridize, leading to a shift of their energies (and hence spectral weight) away from zero, as shown in Fig.~\ref{fig:Fig8} (b). Once the last spin is rotated into the plane, the MZMs are fully hybridized with their spectral weight being completely transferred into the bulk band, as shown in Fig.~\ref{fig:Fig8} (c)  (the full time dependence of this process is shown in Supplemental Movie 1 of the Supplemental Material \cite{supp}). In Fig.~\ref{fig:Fig8} (d), we present the time-dependent difference in the charge density  between the states $|1\rangle$ and $|0\rangle$, i.e., $\Delta \rho_{1,0}({\bf r},t)$ (see Eq.(\ref{eq:rho_diff})). When the fusion process is nearly completed, and only a single out-of-plane spin remains (as indicated by the horizontal dotted white line),  $\Delta \rho_{1,0}({\bf r},t)$ is the largest, and strongly localized at the site of this single out-of-plane spin. However, after the last spin is rotated into the plane, the excess charge delocalizes along the entire chain due to the translational invariance in the interior of the chain. In addition, in Fig.~\ref{fig:Fig8} (e) we present the time dependence of the total  charge difference, $\Delta q_{1,0}(t)$: it begins to deviate from zero once the two MZMs start to hybridize (when the system possesses approximately 17 ferromagnetically aligned spins), and reaches its final value when the system possess a single out-of-plane spin. While the charge delocalizes across the entire spin chain after the last spin is rotated into the plane, the total charge difference remains constant. Interestingly enough, however, it deviates from the expected value of $-e$, as charge (or in general the particle number) is not a good quantum number in the superconducting state due to particle-hole mixing. The deviation from the quantized value of the excess charge of $-e$ is thus expected to increase with increasing magnitude of the superconducting order parameter, as confirmed by our numerical results shown in 
Fig.~\ref{fig:Fig8} (f).

\begin{figure}[h!]
  \centering
  \includegraphics[width=0.95\columnwidth]{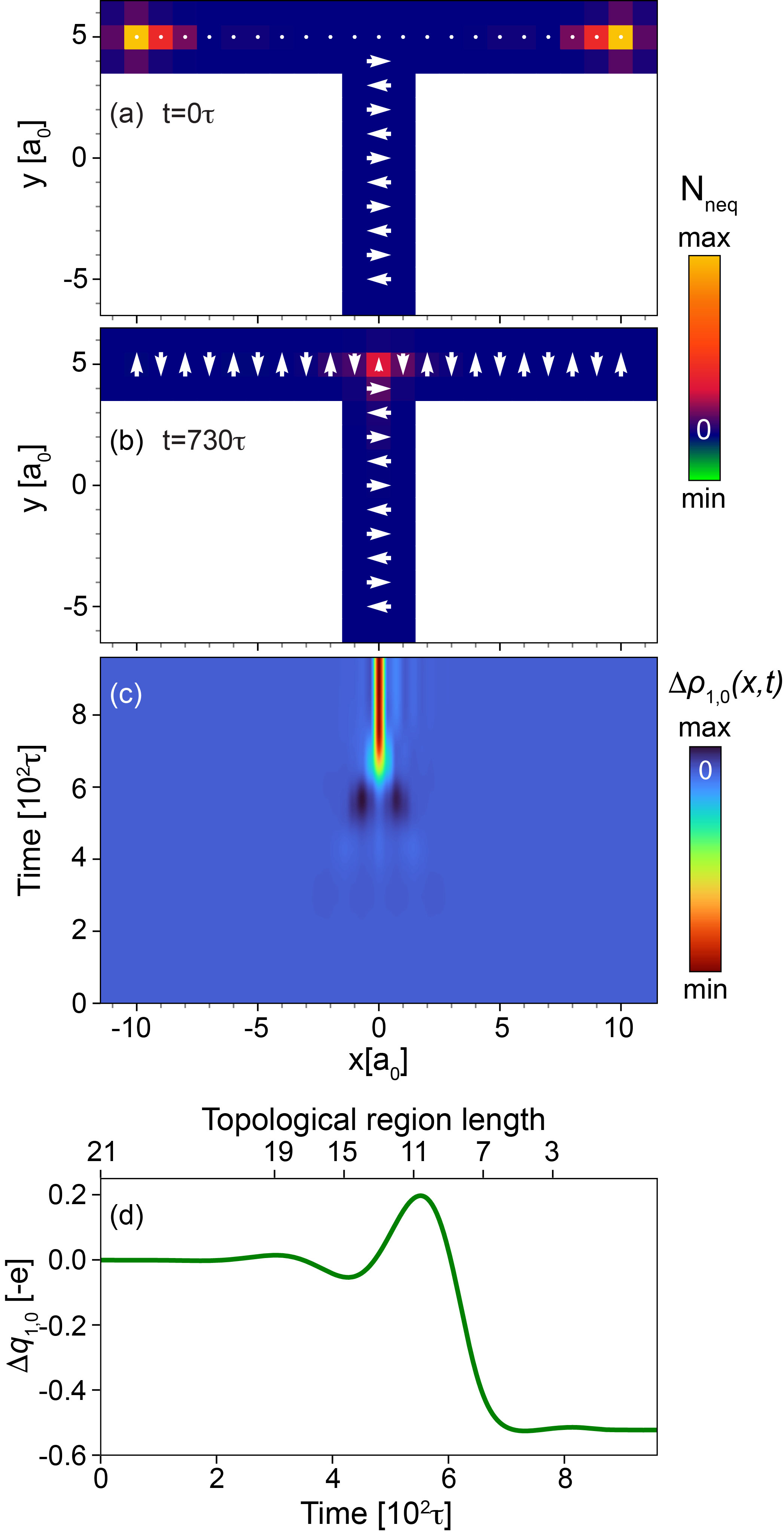}
  \caption{
    Zero-energy $N_\mathrm{neq}$ (a) at the beginning of the fusion process at $t= 0$, and (b) when the MZMs are nearly completely fused at $t=730 \tau$.  The spin orientations are shown as white arrows with white dots indicating an out-of-plane orientaton.
  (c) Time-dependent charge density difference in the horizontal part of the T-structure as a function of position and time.  (d) Time-dependent total charge difference $\Delta q_{1,0}(t)$. The top axis in (d) shows the length of the topological region, i.e.,  the number of out-of-plane aligned spins.
 Chosen parameters are $(\mu, \alpha, \Delta, JS) = (-3.97, 0.9, 2.4, 5.2)\tilde{t}$ with the rotation time $t_R = 300 \tau$ for one spin and the delay time $\Delta t_R = 0.2 t_R$ until the rotation of the next spin.}
  \label{fig:Fig9}
\end{figure}

%
\subsection{MSH T-structure}
%

T-structures, such as the one shown in Fig.~\ref{fig:Fig9} (a) have been proposed as one of the possible architectures to implement quantum gates 
\cite{Alicea2011,Amorim2015,Sanno2021,Mascot2023,Bedow2024,Hodge2024}. 
Hence, being able to read out the result of a gate operation in this geometry is crucial. Figs.~\ref{fig:Fig9} (a) and (b) show the spatial form of the zero-energy $N_\mathrm{neq}$ at the the start of the fusion process at $t=0$ and shortly before its completion at $t=t_f$, respectively, with the fusion occurring at the trijunction point of the T-structure. In Fig.~\ref{fig:Fig9} (c), we present the charge density difference, $\rho_{1,0}({\bf r},t)$ projected onto the $x$-axis, between the $|1\rangle$ and $|0\rangle$ states. While the fusion of the MZMs again leads to a non-zero charge density difference,  we now find that in contrast to the 1D MSH chain discussed above, the charge remains localized near the fusion point for $t > t_f$, as the latter possesses a localized bound state that traps the charge. This trapping of the charge after fusion will likely render its experimental detections via slower charge sensing probes more feasible \cite{Zhou2022}. Moreover, in Fig.~\ref{fig:Fig9} (d), we present the time dependence of the total charge difference, $\Delta q_{1,0}(t)$, which again remains unchanged after the fusion process is concluded at $t=t_f$. Moreover, in contrast to the result from Fig.~\ref{fig:Fig8}, the total charge difference for this case also has the opposite sign, which stems from a crossing of the MZM energy levels through zero energy as they are fused. This results in the breaking of a Cooper pair and a local quantum phase transition during that process, and hence a change in the sign of the excess charge difference.

%
\subsection{Square-and-arms MSH structure }
\subsubsection{Fusion after the execution of an $X$-gate}
%

\begin{figure*}[t]
  \centering
  \includegraphics[width=17cm]{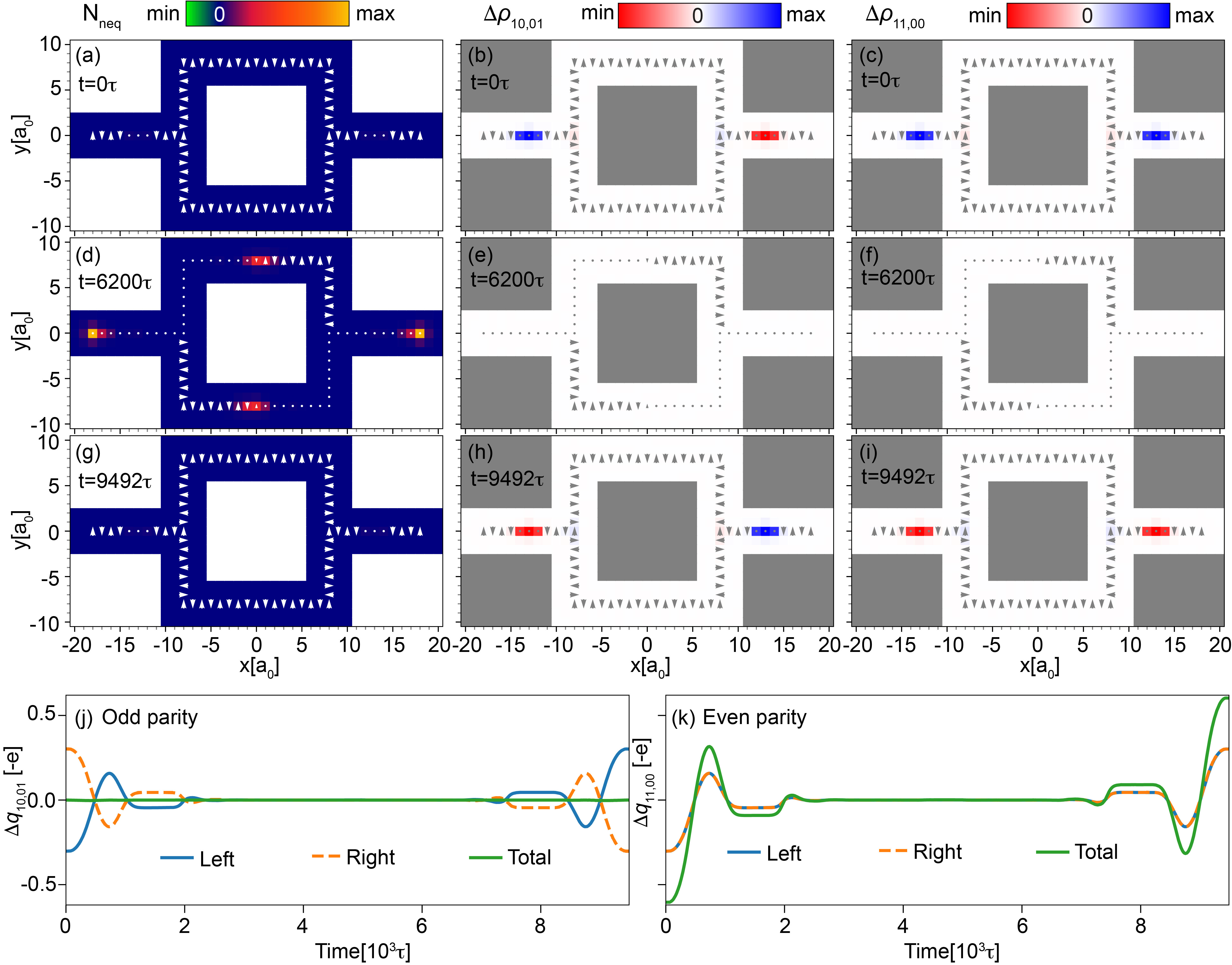}
  \caption{  Zero-energy $N_\mathrm{neq}$ (first column) and and charge density differences $\Delta\rho_{10,01}({\bf r},t)$ (second column) and $\Delta\rho_{11,00}({\bf r},t)$ (third column) for several times during the fusion and X-gate process: (a)-(c) at $t = 0 \tau$ when the MZMs are partially fused, (d)-(f) during the execution of the $X$-gate at $t = 6200 \tau$, and (g)-(i) at $t = 9492 \tau$ after the execution of the $X$-gate is completed and the MZMs are partly fused again. Time-dependent charge difference in the (j) odd parity sector, $\Delta q_{10,01}(t)$, and (k)  even parity sector, $\Delta q_{11,00}(t)$ for the entire system (green line), as well as for the left (blue line) and right (dashed orange line)) halves of the system. Parameters are $(\mu, \alpha, \Delta, J) = (-3.97, 0.9, 2.4, 5.2)\tilde{t}$, with the rotation time $t^f_{R} = 1000 \tau$ for one spin during the fusion process and $t^b_{R} = 430 \tau$ during the braiding process and the delay time $\Delta t_R^{\alpha} = 0.2 t_R^{\alpha}$ until the rotation of the next spin, where $\alpha =f, b$. 
}
  \label{fig:Fig10}
\end{figure*}

For the execution of quantum gates using MZMs, the fusion of MZMs after the completion of the gate process is a crucial step to read-out the final quantum state and evaluate the success of the gate operation. To demonstrate this, we consider $\sqrt{X}$- and $X$-gates, for which two MZMs from different MZM pairs are exchanged once or twice, respectively \cite{Beenakker2020}. As demonstrated in Ref.~\cite{Bedow2024}, a suitable architecture for the implementation of these gates is the square-and-arms architecture, as it allows the use of spatial symmetry to avoid (in the odd parity sector) detrimental effects arising from the hybridization of the MZMs. 

Starting point at $t=0$ is a configuration in which two pairs of MZMs are located in the left and right arms of the square-and-arms structure. The initial states of the system are characterized by $\ket{ij}$ with $i,j=0,1$ in Fock notation, depending on the whether the fusion of the individual MZM pairs leads to the vacuum ($i=0$) or an electron ($i=1$). Thus we have two many-body states in the even parity sector, $\ket{00}$ and $\ket{11}$, which in the logical qubit notation are denoted by $\ket{0}_{\rm logic}$ and $\ket{1}_{\rm logic}$, respectively, and two many-body states in the odd parity sector, $\ket{01}$ and $\ket{10}$, which in the logical qubit notation are denoted by $\ket{0}_{\rm logic}$ and $\ket{1}_{\rm logic}$, respectively. 
The application of the  $X$-gate then yields $X\ket{0}_{\rm logic} = \ket{1}_{\rm logic}$ with the $X$-gate being executed by braiding the two innermost MZMs (belonging to different pairs) twice using the central square, as described in Ref.\,\cite{Bedow2024}. To demonstrate the effect of the $X$-gate on the system, we identify the final many-body states via the spatial structure of its charge density after partial fusion of the MZM pairs.

Since at $t=0$, the topological region in both arms (consisting of only three out-of-plane aligned spins) is shorter than the MZM localization length, the MZMs hybridize and are thus partly fused, shifting their energy away from zero. As a result, the MZMs are not visible in the zero-energy $N_\mathrm{neq}$ shown in Fig.~\ref{fig:Fig10}(a). At the same time, the partial fusion leads to a non-zero charge density difference between the $\ket{10}$ and $\ket{01}$ states, $\Delta \rho_{10,01}({\bf r},t)$ (see Fig.~\ref{fig:Fig10}(b)): at $t=0$, $\Delta \rho$ is positive (negative) on the left (right) arm of the structure, indicating the spatially reversed charge density of the $\ket{10}$ and $\ket{01}$ states. Similarly at $t=0$, there also exists a non-zero difference between the charge densities of the $\ket{11}$ and $\ket{00}$ states, $\Delta \rho_{11,00}({\bf r},t)$ (see Fig.~\ref{fig:Fig10}(c)). As the distance between the MZMs is increased (leading to a significantly reduced hybridization between the MZMs) their energies  shift to zero-energy, and they become visible in the zero-energy $N_\mathrm{neq}$ (see Fig.~\ref{fig:Fig10}(d)). At the same time, $\Delta \rho_{10,01}$ and $\Delta \rho_{11,00}$ vanish, as shown in Figs.~\ref{fig:Fig10}(e) and (f), respectively, as the MZMs are no longer fused.

After the completion of the $X$-gate, and when the MZMs are partly fused again at $t=t_f$ (see Fig.~\ref{fig:Fig10}(g)), the energies of the MZMs are again shifted away from zero, such that the MZMs are not any longer visible in the zero-energy $N_\mathrm{neq}$. At the same time, we find that the differences in the charge densities are reversed from their values at $t=0$, i.e.,  $\Delta \rho_{10,01}({\bf r},t_f) = - \Delta \rho_{10,01}({\bf r},0)$ (cf. Figs.\,\ref{fig:Fig10}(b) and (h)) and $\Delta \rho_{11,00}({\bf r},t_f) = - \Delta \rho_{11,00}({\bf r},0)$ (cf. Figs.~\ref{fig:Fig10}(c) and (i)). This reversal indicates that the $X$-gate has been successfully executed, since $X\ket{0}_{\rm logic} = \ket{1}_{\rm logic}$ and $X\ket{1}_{\rm logic} = \ket{0}_{\rm logic}$, thus reversing the charge density differences, as shown in Fig.\,\ref{fig:Fig10}.

This inversion of the charge density difference is also reflected in the total time-dependent charge difference obtained by integrating the charge density over the left side (solid blue line) or right side (dashed orange line) of the system, as shown in Figs.\,\ref{fig:Fig10}\,(j) and (k) for the odd and even parity sectors, respectively. At the same time, the total charge difference, obtained by interating over the entire system, remains zero in the odd parity sector (as the charge is only exchanged between the left and right arms), but changes in the even parity sector (as a Cooper pair is broken). This inversion, or the ‘magic jump’ \cite{Beenakker2020} of the charge, is one of the hallmarks of MZMs, reflecting their non-Abelian properties.

%
\subsubsection{Fusion after the execution of $\sqrt{X}$ and Hadamard gates}
%

\begin{figure}[t!]
  \centering
  \includegraphics[width=\columnwidth]{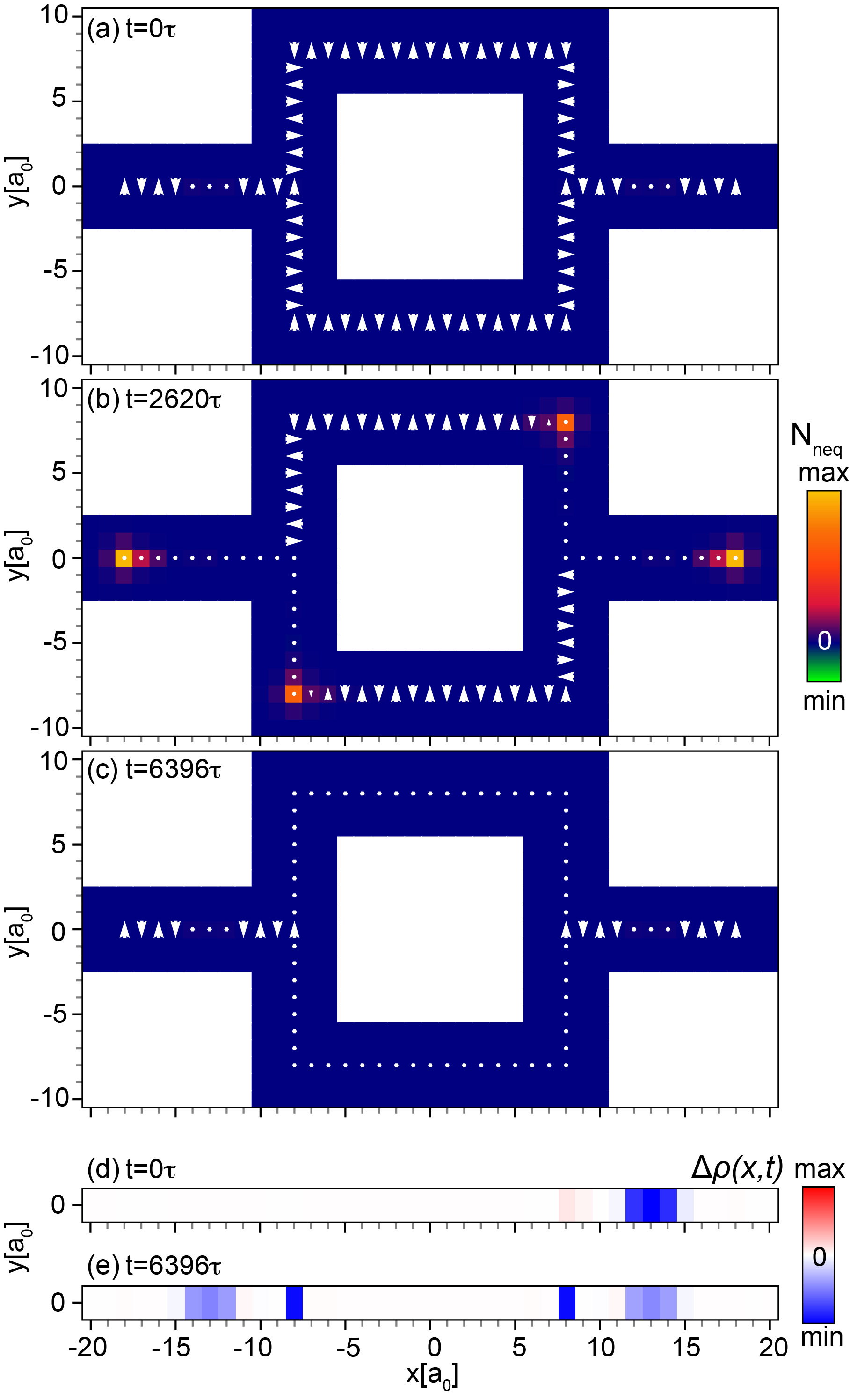}
  \caption{ 
  Zero-energy $N_\mathrm{neq}$, for several times during fusion process and the execution of the  $\sqrt{X}$-gate: (a) at $t = 0$ with the MZMs partially fused, (b) at $t = 2620 \tau$ during the execution of the $\sqrt{X}$-gate, and (c) at $t = 6396 \tau = \tau_f$ after the completion of the $\sqrt{X}$-gate and the partial fusion of the MZMs.
  Charge density difference $\Delta \rho ({\bf r},t)$ along $y=0$ at (d) $t = 0$, and (e) $t = \tau_f$ .
  Parameters are $(\mu, \alpha, \Delta, J) = (-3.97, 0.9, 2.4, 5.2)\tilde{t}$, with the rotation time $t_R^{f} = 1000 \tau$ for one spin for the fusion process and $t_R^b = 430 \tau$ for the braiding process and the delay time $\Delta t_R^{\alpha} = 0.2 t_R^{\alpha}$ until the rotation of the next spin, where $\alpha =f, b$.
}
  \label{fig:Fig11}
\end{figure}

To contrast the cases considered above where the MSH system before fusion is always in a pure qubit state, we consider next the fusion process after the execution of a $\sqrt{X}$-gate, which leaves the system in a mixed qubit state. Preparing the system in the odd-parity state $\ket{10}$, we obtain after the execution of the 
$\sqrt{X}$-gate $\sqrt{X}\ket{01}=\alpha \ket{01} + \beta \ket{10}$ with $|\alpha|^2 = |\beta|^2=\frac{1}{2}$. To investigate the resulting charge density after fusion,  we begin from the same intial spin configuration as for the $X$-gate where the MZMs are partly fused, as shown in Fig.~\ref{fig:Fig11}(a). The $\sqrt{X}$-gate is then performed by rotating the two inner MZMs by $180^\circ$, i.e., exchanging them once (see Fig.~\ref{fig:Fig11}(b)). This is followed by a partial fusion,  resulting in a spin configuration (see Fig.~\ref{fig:Fig11}(c)) that is qualitatively different from the initial one: while the spin orientation in the arms is the same as that at $t=0$, the spins in the square are now aligned ferromagnetically out-of-plane. This leaves the entire box in a topological phase, tough with no additional MZMs as the 
only end points are the junction points with the arms, where the one in-plane magnetic moment is a sufficiently small trivial region that no additional MZMs can emerge.

To investigate the changes in the charge density upon completion of the  $\sqrt{X}$ gate, we compute the time-dependent charge density difference between the initialized state $\ket{01}$, upon which the $\sqrt{X}$ gate is executed, and the state $\ket{00}$ as a reference state, i.e., $\Delta \rho ({\bf r},t) = \rho_{01} ({\bf r},t) - \rho_{00} ({\bf r})$.
At $t=0$, $\Delta \rho ({\bf r},t=0)$ exhibits an excess charge in the right arm (see Fig.\,\ref{fig:Fig11}(d)), where we present the charge density difference at $t=0$ along the center of the system, i.e., for $y=0$) as in the state $\ket{01}$, the MZM pair in the right arm is occupied. After the execution of the $\sqrt{X}$-gate and partial fusion at $t=t_f$, we have 
\begin{align}
    \Delta \rho ({\bf r},t_f) &= \rho_{01} ({\bf r},t_f) - \rho_{00} ({\bf r}) \nonumber \\
    &= \frac{1}{2} \rho_{01} ({\bf r}) + \frac{1}{2} \rho_{10} ({\bf r}) - \rho_{00} ({\bf r})
\end{align}
such that the system now exhibits an equal charge density difference in both arms, as shown in Fig.\,\ref{fig:Fig11}(e). Thus, the excess charge in each of the arms at $t=t_f$ is half of the initial excess charge in the right arm at $t=0$.
Since the final state of the system is a mixed state of the qubit, the final charge distribution should be interpreted probabilistically as an average value: each time the charge is measured experimentally, one either finds an excess charge only in the right or the left arm of the system, and only on average (i.e., by repeating the initialization, $\sqrt{X}$-gate, fusion, and measurement many times) do we obtain the equal charge distribution in both arms shown in Fig.~\ref{fig:Fig11}(e).

%
\section{Discussion}\label{sec:discussion}
%
As well known from other topological systems such as topological insulators, also for the topological superconductor models considered in this paper the gap size determines how localized the edge states are.
It is well-established that a smaller (topological) bulk gap, leading to less localized Majorana edge modes, must be compensated by a larger system size to reduce any possible overlap between the Majorana wave functions at different edges. As a consequence, one can use close-to-realistic parameters and system sizes of several thousands of lattice sites, or one uses significantly larger parameters (i.e., unrealistically large) and reduces the system size to tens of lattice sites per spatial direction. The physical results are usually comparable and the full phenomenology of topological phases can be observed and understood already in small systems. This trade-off between system size and topological gap size is well known. In this paper, we have to overcome the additional challenge that the full lattice system has to be quantum-mechanically time-evolved (for braiding and performing quantum gates): these requirements put a natural upper limit on the system sizes we can compute on state-of-the-art high-performance computing facilities. In order to guarantee sufficiently localized MZMs, the  large values of the superconducting order parameter as used in the manuscript were thus unavoidable in both the Kitaev wire model and in the MSH model. For the same reasons, also the Rashba spin-orbit coupling $\alpha$ was chosen larger than expected in realistic situations. Nevertheless and as stated before, we expect that our results about fusion will quantitatively remain unchanged, even if we used more realistic parameters on much larger system sizes.

%
\section{Conclusion}\label{sec:conclusion}
%

Fusion is a crucial part of any Majorana-based topological quantum computer. 
While fusion has been explored previously in static situations\,\cite{Bai2024, Tsintzis2024, Sau2024, Steiner2020,Boross2024,Verlinde1988, Lahtinen2017, Maciazek2024, Souto2022, Pandey2023, Pandey2023_2, Wang2024}, this paper complements the literature by explicitly performing dynamic simulations of the most important fusion scenarios utilizing a time-dependent methodology to do this on any MZM-based platform.
Our work not only demonstrates $Z$ and $X$ fusion for chains or networks governed by the spinless Kitaev chain model, the minimal model of any one-dimensional topological superconductor. We also investigate fusion in a spinful model describing MSH structures. For both spinless and spinful models, our dynamic simulations confirm the fusion rules of the Ising anyon model. 
We stress that, while quantitative predictions \emph{must} be tailored to specific experimental set-ups (e.g., system size, geometry, parameter set), in essence, our results remain \emph{qualitatively} equivalent as the low-energy physics of MZM-systems are identical, with changes in parameter set simply scaling the topological gap, along with the localization length of the MZMs.
In all, our results constitute a significant step on the way towards the detection of MZMs and their anyonic statistics in real systems, important for the realization of topological quantum computation.

The data for this paper is openly available in Zenodo\,\cite{zenodo}.

\begin{acknowledgments}
S.R.\ acknowledges support from the Australian Research Council through Grants No.\ DP200101118 and No.\ DP240100168.
T.K., J.B, and D.K.M. acknowledge support by the U.\ S.\ Department of Energy, Office of Science, Basic Energy Sciences, under Award No.\ DE-FG02-05ER46225. 
 This research was undertaken using resources from the National Computational Infrastructure (NCI Australia), an NCRIS enabled capability supported by the Australian Government.
\end{acknowledgments}


\appendix

%
\section{A1. Ising Anyon Fusion Algebra Review } \label{sec:Appendix}
%
We give a brief review of key results relating to the fusion space and associative algebra of the SU(2)$_2$ Ising anyon model \cite{Kitaev2006,Nayak2008,Lahtinen2017}, utilized in this paper, and the wider context of TQC as a whole.

We consider a set of particles, or \emph{anyonic charges}, $M \in \{1,a,b,....,\}$, where each element of $M$ denote the \emph{topological charge} associated with each species. 
Over the course of this paper, $1$ will always represent trivial vacuum. 
Fundamental to the algebraic theory of these models is the principle that any two anyonic charges may themselves be combined to form a different anyonic charge within the space, a process known as \emph{fusion}. 
Such a fusion process may between two such anyonic charges, $a,b\in M$, is denoted by  
\begin{equation}
    a \times b=\sum_c N^c_{ab}c.
\end{equation}
Here, $N^c_{ab}\in \mathbb{Z}_{\geq 0}$ gives the number of possible fusion channels available for $a$ and $b$ to fuse to anyon $c$.

Importantly, we may associate each fusion channel with a state vector, $|ab;c\rangle$.
We associate the fusion process associated with this state vector diagramatically via use of a \emph{fusion tree}, given by
\begin{equation}
    |ab;c\rangle= \begin{tikzpicture}[baseline=-2.5ex]
\node (a) at (0,-0.0)
 {
\begin{tikzpicture}
  [
  line width=0.25mm, 
  level distance=7.5mm,
   level 1/.style={sibling distance=20mm},
   level 2/.style={sibling distance=10.3mm},
   grow'=up]
  \coordinate
  child {edge from parent } {
    child child
     };
\node [anchor=north,yshift=1.95cm,xshift=-1.1cm] {$a$};
\node [anchor=north,yshift=1.95cm,xshift=1.1cm] {$b$};
\node [anchor=north,yshift=0cm,xshift=0.0cm] {$c$};
\end{tikzpicture}};
\end{tikzpicture}.
\end{equation}
We define the set $V^{ab}_c=\{|ab;c\rangle\}$, where we call $V^{ab}_c$ the \emph{fusion space}.

For a system containing more than one anyon, $M=\{a_1,a_2,....,a_n\}$, we may define the fusion space $V^{a_1,a_2,...,a_n}_d=\bigoplus_{c_1,c_2,...,c_{n-1}}V^{a_1,a_2}_{c_1}\otimes V^{c_1,a_3}_{c_2} \otimes ... \otimes V^{c_{n-1},a_n}_{d}$. 
While $d$ is fixed, conserving the total topological charge in the system, $c_1,c_2,... c_{n-1}\in M$ are free to take on different configurations, leading to many possible states $|a_1a_2;c_1a_3;c_2a_4;...;c_{n-1}a_n;d\rangle \in V^{a_1,a_2,...,a_n}_d$ in the fusion space.




For the purpose of this paper, we represent the fusion space in the basis whereby each neighbouring anyon is pairwaise fused, which we denote as the \emph{Z-basis}. 
For an $n$-anyon system, we construct the Hilbert space $V^{Z_1Z_2...Z_n}_g$ such that $V_g^{Z_1Z_2...Z_n}=\{|(a_1a_2)(a_3a_4)...;(n_1n_2)...;g\rangle\}_{a_i \in M}$, where $(a_ia_j)$ corresponds to the pairwise fusion of anyonic charges $a_i$, $a_j$.
In the case of a four anyon initialization with fixed total topological charge $g$, $V_g^{Z_1Z_2}$, we associate an element of the fusion space with a set state vector $\{|(ab)(cd);ef;g\rangle\}$.  
We may diagramatically associate this vector with a \emph{fusion tree}, such that:
\begin{equation}
    |(ab)(cd);ef;g\rangle
    =\begin{tikzpicture}[baseline=-1.2ex]
\node (a) at (0,-0.3)
 {
\begin{tikzpicture}
  [
  line width=0.25mm, 
  level distance=7.5mm,
   level 1/.style={sibling distance=20mm},
   level 2/.style={sibling distance=10.3mm},
   grow'=up]
  \coordinate
  child {edge from parent } {
    child {
       child
       child
     }
     child {
       child 
       child 
     }};
\node [anchor=north,yshift=2.7cm,xshift=-1.52cm] {$a$};
\node [anchor=north,yshift=2.7cm,xshift=-0.47cm] {$b$};
\node [anchor=north,yshift=2.7cm,xshift=0.47cm] {$c$};
\node [anchor=north,yshift=2.7cm,xshift=1.52cm] {$d$};
\node [anchor=north,yshift=1.2cm,xshift=-0.7cm] {$e$};
\node [anchor=north,yshift=1.2cm,xshift=0.7cm] {$f$};
\node [anchor=north,yshift=0cm,xshift=0cm] {$g$};
\end{tikzpicture}
};
\end{tikzpicture}.
\end{equation}

Here, each point vertex corresponds to a fusion of two species into a third species such species. 
We may then fuse each element in the order given until the there is only one topological charge, $g$, left in the space. 
This provides a well defined basis set for our fusion space. 

It is possible to change the fusion order by a linear transformation $F$, otherwise known as an \emph{F-move}. 
With respect to a state $|(ab)(cd);ef;g\rangle$ in the initial fusion space $V$, an F-move will corresponds to the following transformation:
\begin{equation}
\begin{tikzpicture}[baseline=-1.2ex]
\node (a) at (0,-0.3)
 {
\begin{tikzpicture}
  [
  line width=0.25mm, 
  level distance=7.5mm,
   level 1/.style={sibling distance=15mm},
   level 2/.style={sibling distance=10.3mm},
   grow'=up]
  \coordinate
  child {edge from parent } {
    child {
       child
       child
     }
     child {
       child 
       child 
     }};
\node [anchor=north,yshift=2.7cm,xshift=-1.25cm] {$a$};
\node [anchor=north,yshift=2.7cm,xshift=-0.24cm] {$b$};
\node [anchor=north,yshift=2.7cm,xshift=0.24cm] {$c$};
\node [anchor=north,yshift=2.7cm,xshift=1.25cm] {$d$};
\node [anchor=north,yshift=1.2cm,xshift=-0.7cm] {$e$};
\node [anchor=north,yshift=1.2cm,xshift=0.7cm] {$f$};
\node [anchor=north,yshift=0cm,xshift=0cm] {$g$};
\end{tikzpicture}
};
\end{tikzpicture}
=\sum_h\left(F^g_{abf}\right)_{eh}
        \begin{tikzpicture}[baseline=-1.2ex]
\node (a) at (0,-0.3)
 {
\begin{tikzpicture}
  [line width=0.25mm, 
  level distance=6.0mm,
   level 1/.style={sibling distance=8mm},
   level 2/.style={sibling distance=8.3mm},
   grow'=up]
  \coordinate
  child {edge from parent } {
  child [grow=130] {child [grow=130] {child [grow=130] {child [grow=130]}}}
    child  {child { child [grow=125] }
       child {child child} }  
     
     };
\node [anchor=north,yshift=2.8cm,xshift=-1.62cm] {$a$};
\node [anchor=north,yshift=2.85cm,xshift=-0.47cm] {$b$};
\node [anchor=north,yshift=2.8cm,xshift=0.3cm] {$c$};
\node [anchor=north,yshift=2.85cm,xshift=1.32cm] {$d$};
\node [anchor=north,yshift=1.7cm,xshift=1.05cm] {$f$};
\node [anchor=north,yshift=1.0cm,xshift=0.5cm] {$h$};
\node [anchor=north,yshift=0cm,xshift=0cm] {$g$};
\end{tikzpicture}
};
\end{tikzpicture}
\end{equation}
providing an isomorphic mapping between different fusion spaces \cite{Kitaev2006}. 
We define the \emph{X-basis}, $V^X_g$, to correspond to the space whereby $b$ and $c$ will be the first pair of anyons to fuse. 
We can represent this diagramatically as 
\begin{equation}
    |a(bc)d;a(ed);af;g\rangle=
    \begin{tikzpicture}[baseline=-1.2ex]
\node (a) at (0,-0.3)
 {
\begin{tikzpicture}
  [line width=0.25mm, 
  level distance=6.0mm,
   level 1/.style={sibling distance=12mm},
   level 2/.style={sibling distance=8.3mm},
   grow'=up]
  \coordinate
  child {edge from parent } {
   child [grow=130] {child [grow=130] {child [grow=130] {child [grow=130]}}}
    child  {child { child           child}
       child [grow=45]{child[grow=45] {child[grow=45]}}   }
     };
\node [anchor=north,yshift=2.8cm,xshift=-1.72cm] {$a$};
\node [anchor=north,yshift=2.9cm,xshift=-0.35cm] {$b$};
\node [anchor=north,yshift=2.9cm,xshift=0.7cm] {$c$};
\node [anchor=north,yshift=2.9cm,xshift=1.92cm] {$d$};
\node [anchor=north,yshift=1.6cm,xshift=0.2cm] {$e$};
\node [anchor=north,yshift=1.0cm,xshift=0.46cm] {$f$};
\node [anchor=north,yshift=0cm,xshift=0cm] {$g$};
\end{tikzpicture}
};
\end{tikzpicture}
\end{equation}
where $|a(bc)d;a(ed);af;g\rangle \in V^X_g$. 
Importantly, we may map between the Z and X-basis by the mapping $F_{Z \to X}$, which, in essence, corresponds to a series of F-moves: $F_{Z \to X}: \; V^{Z_1Z_2}_g \to V^{X}_g,$ $F_{Z \to X}(|(ab)(cd);ef;g\rangle)=\sum_{hi}(F^g_{abf})_{eh}(F^h_{bcd})^{-1}_{fi}|a(bc)d;a(fd);ai;g\rangle$.

The mutual statistics of the anyons in the fusion space under exchange may be encoded with a unitary known as an \emph{R-move}, which corresponds to a clockwise rotation of the anyons. 
For a clockwise exchange between $a$ and $b$, we can denote the action of this transformation as 
\begin{equation}
\begin{tikzpicture}[baseline=-1.8ex]
       \node (a) at (0,0.0)
         {
\begin{tikzpicture}
[
every braid/.style={
ultra thick,
braid/strand 1/.style=black,
braid/strand 2/.style=black,
braid/strand 3/.style=black,
braid/strand 4/.style=black,
braid/anchor=center,
braid/number of strands=4,
braid/height=-0.4cm
}
]
\pic[
 braid/.cd,
 gap=0.25,
 width=1.0cm,
 every strand/.style={line width=0.25mm},
 number of strands=4,
 height=-0.4cm
] {braid={
s_1^{-1}
}};
\end{tikzpicture}
};
\node [anchor=north,yshift=0.9cm,xshift=-1.52cm] {$a$};
\node [anchor=north,yshift=0.9cm,xshift=-0.47cm] {$b$};
\node [anchor=north,yshift=0.9cm,xshift=0.47cm] {$c$};
\node [anchor=north,yshift=0.9cm,xshift=1.52cm] {$d$};
\node [anchor=north,yshift=-1.0cm,xshift=-0.65cm] {$e$};
\node [anchor=north,yshift=-1.0cm,xshift=0.65cm] {$f$};
\node [anchor=north,yshift=-1.8cm,xshift=0cm] {$g$};
\node (b) at (a.south) [anchor=north,yshift=0.31cm,grow'=up]
{
\begin{tikzpicture}
  [line width=0.25mm, 
  level distance=4mm,
   level 1/.style={sibling distance=20mm},
   level 2/.style={sibling distance=10.3mm}]
  \coordinate
  child {edge from parent } {
    child {
       child
       child
     }
     child {
       child 
       child 
     }};
\end{tikzpicture}
};
\end{tikzpicture}
        =R^e_{ab}
\begin{tikzpicture}[baseline=-2.5ex]
\node (a) at (0,-0.7)
 {
\begin{tikzpicture}
  [
  line width=0.25mm, 
  level distance=7.5mm,
   level 1/.style={sibling distance=20mm},
   level 2/.style={sibling distance=10.3mm},
   grow'=up]
  \coordinate
  child {edge from parent } {
    child {
       child
       child
     }
     child {
       child 
       child 
     }};
\node [anchor=north,yshift=2.7cm,xshift=-1.52cm] {$a$};
\node [anchor=north,yshift=2.7cm,xshift=-0.47cm] {$b$};
\node [anchor=north,yshift=2.7cm,xshift=0.47cm] {$c$};
\node [anchor=north,yshift=2.7cm,xshift=1.52cm] {$d$};
\node [anchor=north,yshift=1.2cm,xshift=-0.7cm] {$e$};
\node [anchor=north,yshift=1.2cm,xshift=0.7cm] {$f$};
\node [anchor=north,yshift=0cm,xshift=0cm] {$g$};
\end{tikzpicture}
};
\end{tikzpicture}
\end{equation}

Finally, we encode an exchange of species in the fusion tree in the form of a \emph{braid}: 

\begin{equation}
\begin{tikzpicture}[baseline=-3.8ex]
       \node (a) at (0,0.0)
         {
\begin{tikzpicture}
\pic[
 braid/.cd,
 gap=0.25,
 width=1.0cm,
 every strand/.style={line width=0.25mm},
 number of strands=4,
 height=-0.4cm
] {braid={
s_2^{-1}
}};
\end{tikzpicture}
};
\node [anchor=north,yshift=0.9cm,xshift=-1.52cm] {$a$};
\node [anchor=north,yshift=0.9cm,xshift=-0.47cm] {$b$};
\node [anchor=north,yshift=0.9cm,xshift=0.47cm] {$c$};
\node [anchor=north,yshift=0.9cm,xshift=1.52cm] {$d$};
\node [anchor=north,yshift=-1.2cm,xshift=-0.7cm] {$e$};
\node [anchor=north,yshift=-1.2cm,xshift=0.7cm] {$f$};
\node [anchor=north,yshift=-1.8cm,xshift=0cm] {$g$};
\node (b) at (a.south) [anchor=north,yshift=0.31cm,grow'=up]
{
\begin{tikzpicture}
  [line width=0.25mm, 
  level distance=4mm,
   level 1/.style={sibling distance=20mm},
   level 2/.style={sibling distance=10.3mm}]
  \coordinate
  child {edge from parent } {
    child {
       child
       child
     }
     child {
       child 
       child 
     }};
\end{tikzpicture}
};
\end{tikzpicture}
        =\sum_{hi}(B_{bc})^{ef}_{hi}
\begin{tikzpicture}[baseline=-1.2ex]
\node (a) at (0,-0.3)
 {
\begin{tikzpicture}
  [
  line width=0.25mm, 
  level distance=7.5mm,
   level 1/.style={sibling distance=20mm},
   level 2/.style={sibling distance=10.3mm},
   grow'=up]
  \coordinate
  child {edge from parent } {
    child {
       child
       child
     }
     child {
       child 
       child 
     }};
\node [anchor=north,yshift=2.7cm,xshift=-1.52cm] {$a$};
\node [anchor=north,yshift=2.7cm,xshift=-0.47cm] {$b$};
\node [anchor=north,yshift=2.7cm,xshift=0.47cm] {$c$};
\node [anchor=north,yshift=2.7cm,xshift=1.52cm] {$d$};
\node [anchor=north,yshift=1.2cm,xshift=-0.7cm] {$h$};
\node [anchor=north,yshift=1.2cm,xshift=0.7cm] {$i$};
\node [anchor=north,yshift=0cm,xshift=0cm] {$g$};
\end{tikzpicture}
};
\end{tikzpicture}
\end{equation}
which corresponds to the mapping $|(ac)(bd);ef;g\rangle=\sum_{hi}(B_{bc})^{ef}_{hi}|(ab)(cd);hi;g\rangle$ where total topological charge $g$ must be conserved.
This corresponds to a \emph{clockwise} exchange of anyons $b$ and $c$, and encodes the anyonic exchange statistics of the particles in the fusion space. 
The $B$-matrix itself may be encoded via a combination of $F$ and $R$ moves. 

Both $F$ and $R-$moves must generically satisfy a set of associative relations known as the \emph{Hexagon} and \emph{Pentagon equations} to be well-defined \cite{Kitaev2006,Nayak2008,Maciazek2024}.  
In the case of the Ising anyon model, the fusion rules are given in Eq. \,\eqref{eq:fuserules} of the main text. 
This fixes the forms of both the $F$-matrix and $R$-matrices are well known, and take the following forms:
\begin{equation}
    F^{\sigma}_{\sigma \sigma \sigma}=\frac{1}{\sqrt{2}}\begin{pmatrix}
        1 & 1 \\ 
        1 & -1
    \end{pmatrix}, \quad R_{\sigma\sigma}=e^{-i\frac{\pi}{8}}\begin{pmatrix}
                            1 & 0\\ 
                            0 & i
                        \end{pmatrix} \label{eq:IsingFR}
\end{equation}
with additional elements of $F^a_{bcd},R_{ab}\in U(1)$. 
Up to a constant phase factor, this sets the braid matrix, $B$, to be $B_{\sigma\sigma} \equiv B=F^{-1}RF$, where $F\equiv F^{\sigma}_{\sigma\sigma\sigma}$ and $g$ sets the total topologcial charge.
In this case, we may, up to a global gauge, map between the Z and X-basis by a single F-move, i.e. $F_{Z \to X}(|\psi\rangle) = F|\psi\rangle$.
This provides the motivation for the purpose of quantum computation. 
We may encode a logical qubit in the this fusion space by mapping logical states to certain fusion states.  
In the even parity subspace, where the total topological charge of the fusion space is $0$, this encoding is given by $|0\rangle \equiv |(\sigma\sigma)(\sigma \sigma);11;1\rangle$, $|1\rangle \equiv |(\sigma\sigma)(\sigma\sigma);\zeta\zeta;1\rangle$.
In this case, we may encode a unitary $\sqrt{Z}$ and $\sqrt{X}$ gate by
\begin{equation}
\begin{aligned}
    &\sqrt{Z}\equiv R_{\sigma\sigma} = e^{-i\frac{\pi}{8}}\begin{pmatrix}
            1 & 0\\ 
            0 & i
        \end{pmatrix}, \\
        &\sqrt{X} \equiv B=\frac{e^{i\frac{\pi}{8}}}{\sqrt{2}}\begin{pmatrix}
            1 & -i\\ 
            -i & 1
        \end{pmatrix}.
 \label{eq:Braid}
 \end{aligned}
\end{equation}

This returns the usual unitary Pauli transformations, with other Pauli transformations given by the compositions $Z=(\sqrt{Z})^2$, $X=(\sqrt{X})^2$, $H=\sqrt{X}\sqrt{Z}^{-1}\sqrt{X}$, $\sqrt{Y}=\sqrt{X}^{-1}\sqrt{Z}\sqrt{X}$.
Thus, in this basis, we may map between different states within the fusion space by $R$-moves and braids, with the final state outcome revealed by measurement of the total charge of the fusion outcomes $Q$. 
For $Q=0$, the state is in the $|(\sigma\sigma)(\sigma\sigma);11;1\rangle$, with $Q=-2e$ corresponding to $|(\sigma\sigma)(\sigma\sigma);\zeta\zeta;1\rangle$.
\bibliography{msh_fusion}
\end{document}